\documentclass[aps,prc,letterpaper,showpacs,12pt,floatfix,preprint]{revtex4-2}

\usepackage{natbib}
\usepackage{amsmath}
\usepackage{graphicx}
\usepackage{hyperref}
\usepackage[utf8]{inputenc}
\usepackage{color}
\usepackage{empheq}

\usepackage{bm}

\allowdisplaybreaks

\begin{document}

	\title{Axial-vector Current and General Unpolarized Electroweak Single-nucleon Responses}
	
	\author{T. W. Donnelly$^{(1)}$ and Sabine Jeschonnek$^{(2)}$}
	
	\affiliation{\small \sl 
		(1) Center for Theoretical Physics, Department of Physics and Laboratory for Nuclear Science, Massachusetts Institute of Technology, Cambridge, MA 02139 \\
		(2) The Ohio State University, Physics
		Department, Lima, OH 45804		 
	}

	\date{\today}

	\begin{abstract}
The present study provides an extension to our recent work on the vector (V) electromagnetic single-nucleon current and associated response functions, both for unpolarized situations and in situations where the target nucleon is polarized. Here the axial-vector (A) single-nucleon current matrix element is developed in detail and the full set of vector and axial-vector currents used to obtain the electroweak VV, AA and VA response functions. Only the unpolarized case is studied in the present work. The general forms for all of these elements are developed together with various approximation schemes in which numerical studies are provided to indicate where these approximations may be expected to be valid. The results of this work provide the basis for a deeper understanding of the roles played by the various single-nucleon form factors in weak interaction reactions on free nucleons and when using the standard ``prescription for nuclear physics'' in reactions involving nucleons in nuclei.
	\end{abstract}

	\maketitle


\section{\protect\bigskip Introduction\label{sec-intro}}

The present study builds on the work presented in \cite{JandD,Donnelly:2024hqx} in which the vector (V) electromagnetic single-nucleon current matrix elements and associated response functions were developed. Here the axial-vector (A) single-nucleon current matrix elements are similarly developed and the weak interaction VV, AA and VA response functions for unpolarized situations provided. The scope is limited in that only first-class currents are retained, although relatively straightforward further extensions could relatively easily be undertaken to include second-class currents as well. The general results obtained can be used directly to obtain weak interaction cross sections for any process involving only single nucleons, as well as to provide the basis for weak interaction processes in nuclei where weak interactions with the nucleons in the nuclei form the underlying reaction mechanism. Going beyond these general results, the present work also yields suggestions for various approximations to the current matrix elements and response functions, including numerical studies in the particular case of isovector currents.

Indeed, such procedures for providing an approach to studying electroweak interactions with single nucleons in free space and then embedding the currents so-obtained for bound nucleons in nuclei have been pursued for decades --- see \cite{Walecka,DW,Donnelly:1978tz} for several review articles on this subject where similar notation is employed. The formalism for various low-energy weak interaction processes (beta decay, electron capture and muon capture) have been presented in those references, together with extensions to higher-energy processes, namely, both charge-changing (CC) and neutral current (NC) neutrino and anti-neutrino reactions with nuclei. In recent years these higher-energy reactions have constituted much of the experimental effort in this area at facilities ranging from medium energies {\it e.g.} at Oak Ridge (COHERENT), to somewhat higher energies {\it e.g.} at J-PARC (T2K)  and to much higher, multi-GeV energies at Fermilab, {\it e.g.} MicroBooNE, ICARUS, SBND, MINERvA and DUNE. And, given the very small cross sections that occur,  and the experimental practicalities that are present, most of these inevitably involve nuclear targets; accordingly, for such neutrino and anti-neutrino reactions at energies of a GeV and beyond the issues we consider in the present work are especially relevant.

The paper is organized in the following way: after a brief discussion of the basic kinematics for electroweak processes involving single nucleons in Sect.~\ref{sec-basic}, the single-nucleon axial-vector current is developed in Sect.~\ref{sec-current}, building on the previous studies in \cite{JandD,Donnelly:2024hqx}. As in those previous studies the spinor matrix elements of the various required gamma matrices are explicitly evaluated and the spin structure of the resulting current matrix elements revealed. Moreover, the results are cast in forms that display the general dependence on the 3-momentum $p$ of the struck nucleon in the single-nucleon matrix elements with the goal of indicating how expansions arise for $p \ll m_N$, where $m_N$ is the nucleon mass. These developments are followed in Sect.~\ref{sec-responses} by discussions of the general  VV, AA and VA electroweak response functions, including in two subsections brief treatments of expansions in $p/m_N$ and results for the Breit frame. In Sect.~\ref{sec-results} explicit results are presented for the specific case occurring for charge-changing neutrino and anti-neutrino reactions, namely, for isovector currents (see, for instance, \cite{usual} for other closely related work and references therein). Full results are given, as well  as various approximations (see \cite{Donnelly:2024hqx} for a discussion of the strategy employed). Finally, the paper ends with a brief set of conclusions in Sect.~\ref{sec-conclusions} and an appendix where our conventions are summarized.

\section{\protect\bigskip Basic Kinematics for electroweak processes \label{sec-basic}}

Building on the developments given in \cite{JandD,Donnelly:2024hqx} for electron scattering and using the conventions set forth in the Appendix, in this section we
begin with a brief discussion of the kinematics involved in general electroweak studies of
lepton scattering from nucleons or nuclei. We consider the case where an incident beam of leptons with 4-momentum $K^{\mu} = (\epsilon, {\bm k})$ scatters and a lepton with 4-momentum $K'^{\mu} = (\epsilon', {\bm k}')$ emerges.
One has
\begin{eqnarray}
	\epsilon &=& \sqrt{m^2 + k^2}  \nonumber \\
	\epsilon' &=& \sqrt{m'^2 + k'^2} ,
\end{eqnarray}
where $m$ and $m'$ are the masses of the incident and outgoing leptons, respectively. Clearly,
for electron scattering $m = m' = m_e$. Usually, but not always, this can be taken to be zero. Also, for neutral-current neutrino scattering one may
almost always take the neutrino mass to be zero. For electron neutrino induced charge-changing
neutrino reactions $m = m_{\nu e} \cong 0$, whereas $m' = m_e$ (again, essentially zero). The
case for muon neutrino induced charge-changing neutrino reactions is somewhat different,
since there $m = m_{\nu \mu} \cong 0$, whereas $m' = m_{\mu}$; the last is clearly not negligible for some of the kinematics of interest experimentally. As usual one has 4-momentum transfer $Q^{\mu} = (\omega, {\bm q})$ with
\begin{eqnarray}
	\omega &=&  \epsilon - \epsilon' \nonumber \\
	{\bm q} &=& {\bm k} - {\bm k}' ,
	\label{defomegaq}
\end{eqnarray}
the energy transfer and 3-momentum transfer, respectively, and where the 4-momentum transfer is
spacelike: $-Q^2 = q^2 - \omega^2 > 0$ (see the Appendix for our conventions).

\begin{figure}
	\centering
	\includegraphics[width=16cm]{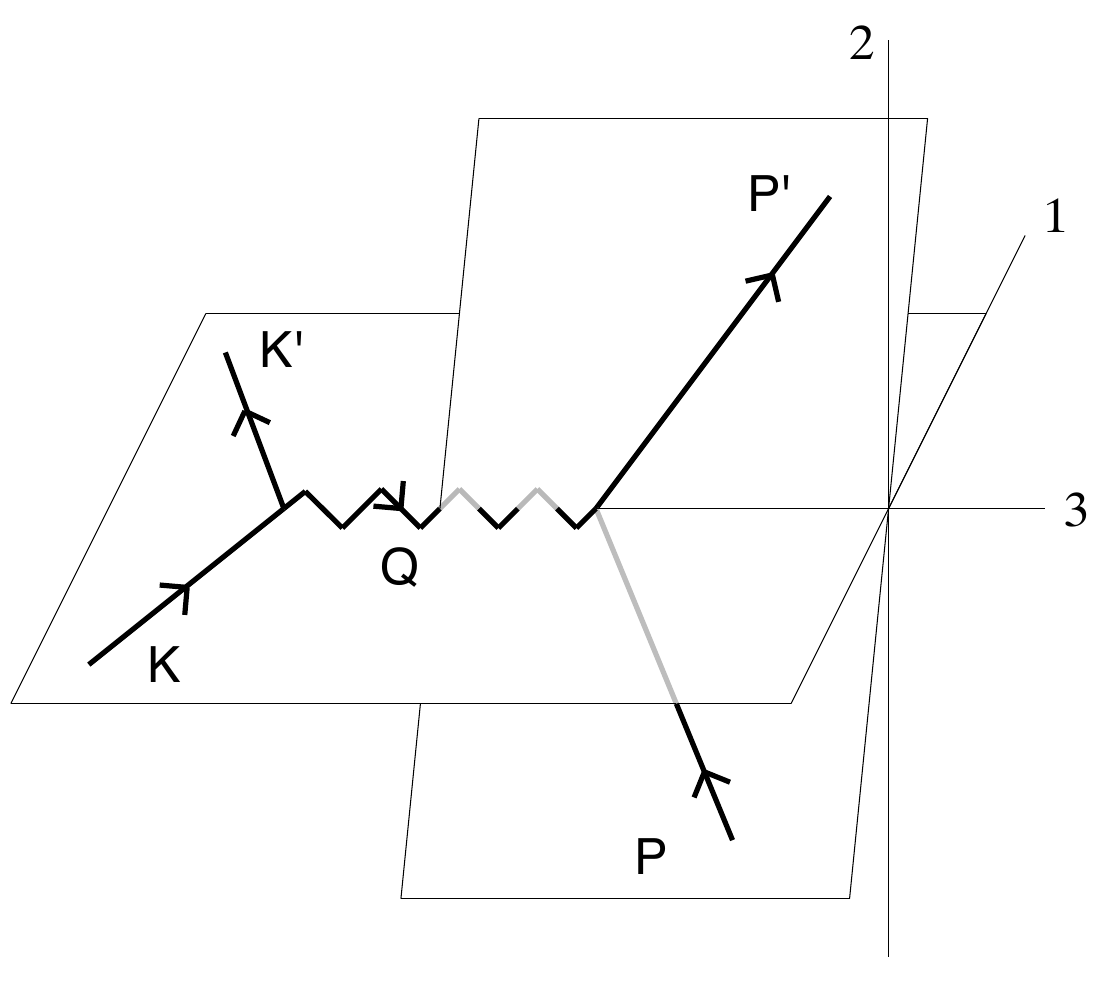}  				
	\caption{Kinematics for charge-changing neutrino or anti-neutrino reactions with single nucleons. Here an incident neutrino or anti-neutrino with 4-momentum $K^\mu$ changes into a negative or positive lepton of the same flavor, respectively, with 4-momentum ${K'}^\mu$ and exchanges a $W^+$ or $W^-$, respectively with 4-momentum $Q^\mu$. Correspondingly, a neutron with 4-momentum $P^\mu$ converts to a proton with 4-momentum ${P'}^\mu$ or {\it vice versa} (see text). Clearly a similar diagram occurs for neutral current processes involving either electrons for studies of parity-violating electron scattering or for neutrinos/anti-neutrinos in NC scattering. }
	\label{fig_kine} 
\end{figure}

The developments above are quite general --- we now make them more specific by assuming,
as is the focus of the present study, that we will limit ourselves to the single-nucleon
electroweak current, namely to the processes shown in Fig.~\ref{fig_kine}. We also limit the discussion to neutral-current (NC) scattering from
protons or neutrons, as well as charge-changing (CC) neutrino and anti-neutrino reactions
with nucleons in both the initial and final states, although the developments may easily
be extended to include low-energy processes such as beta-decay and muon-capture \cite{Walecka,Donnelly:1978tz}, or to parity-violating electron scattering \cite{PVEE}. Let us denote the initial nucleon mass by $M$ and the final nucleon
mass by $M'$, where then the initial and final 4-momenta of the nucleon are $P^{\mu} = (E_p, {\bm p})$
and $P^{\prime \mu} = (E_{p\prime}, {\bm p}')$, respectively, with $E_p = (p^2 +M^2)^{1/2}$ and $E_{p^\prime} = (p^{\prime 2} +M^{\prime 2})^{1/2}$ and from
4-momentum conservation we have
\begin{eqnarray}
	Q^{\mu} &=& P^{\prime \mu} - P^{\mu} \nonumber \\
	\omega &=& E_{p^{\prime}} - E_p \nonumber \\
	{\bm q} &=& {\bm p} ^{\prime} - {\bm p} \, .
\end{eqnarray}
For our choice of coordinate systems we use the conventions discussed in \cite{Donnelly:2024hqx}, specifically for the 123-system we have that the leptons lie in the ${\mathbf u}_1 - {\mathbf u}_3$ plane with ${\mathbf q}$ along ${\mathbf u}_3$. The 3-momentum ${\mathbf p}$ is given in terms of the magnitude of its momentum $p=|{\mathbf p}|$ and the angles $\theta$ and $\phi$  by
\begin{equation}
{\mathbf p} \equiv p \left[ \sin\theta \cos\phi {\mathbf u}_1 + \sin\theta\sin\phi {\mathbf u}_2 + \cos\theta {\mathbf u}_3 \right] ,\label{eq-p}
\end{equation}
while the $1'2'3'$-system is defined by rotating through $\phi$ so that the 3-vectors ${\mathbf q}$ and ${\mathbf p}$ lie in the ${\mathbf u}'_1 - {\mathbf u}'_3$ plane.

For the neutrino reactions
one has the following: the hadronic masses above are those of the nucleon, namely,
$M = W = M_p$ or $M = W = M_n$ for NC scattering from protons or neutrons, respectively, namely $\nu_\ell + p \rightarrow \nu_\ell + p$ or $\nu_\ell + n \rightarrow \nu_\ell + n$ and ${\bar\nu}_\ell + p \rightarrow {\bar\nu}_\ell + p$ or ${\bar\nu}_\ell + n \rightarrow {\bar\nu}_\ell + n$, where $\ell = e$, $\mu$ or $\tau$. Moreover, one has $M = M_n$ and $W = M_p$ for CC neutrino reactions, $\nu_\ell + n \rightarrow \ell^- + p$ and $M = M_p$ and $W = M_n$ for CC
anti-neutrino reactions, ${\bar\nu}_\ell + p \rightarrow \ell^+ + n$. Later we shall limit the discussion to CC neutrino and anti-neutrino reactions, although similar developments may be undertaken for other related electroweak processes (see, for example, \cite{Donnelly:1978tz}).

We then make the commonly adopted definition of a universal mass for the nucleon, namely,
\begin{equation}
	m_N \equiv (M_p + M_n)/2  \,.
\end{equation}
and then may employ the same kinematic variables as in studies of elastic electron scattering \cite{JandD,Donnelly:2024hqx}. In those references we employ dimensionless variables:
\begin{eqnarray}
	\lambda &\equiv& \omega/2 m_N \nonumber \\
	{\bm \kappa} &\equiv& {\mathbf q}/2 m_N \nonumber \\
	\tau &\equiv& \frac{|Q^2|}{4 m_N^2}
	= \kappa^2 -\lambda^2 \ge 0 \nonumber \\
	{\bm \eta} &\equiv& {\mathbf p}/m_N \nonumber \\
	\varepsilon &\equiv& E_p/m_N = \sqrt{1+\eta^2}  \nonumber \\
	\delta &\equiv& \eta \sin \theta \nonumber \\
	\delta^\prime &\equiv& \eta \cos \theta  \,, \label{eq-sn-74}
\end{eqnarray}
where $\theta$ is the angle between ${\mathbf q}$ and ${\mathbf p}$. From energy-momentum conservation we have that
\begin{eqnarray}
	\kappa \delta^\prime &=& \lambda \varepsilon - \tau + {\tilde \delta} = \lambda {\bar \varepsilon} - \kappa^2 + {\tilde \delta}  \nonumber \\
	{\bar \varepsilon} &\equiv& \varepsilon + \lambda =  \frac{1}{\sqrt{\tau}} \left[ \kappa^2 \left( 1 + \tau + \delta^2 \right) + {\tilde \delta} \left( 2 \kappa \delta' - {\tilde \delta} \right) \right]^{1/2 } , \label{eq-sn-77}
\end{eqnarray}
where
\begin{equation}
{\tilde \delta} \equiv \frac{M-M'}{M+M'} \cong \pm7\times 10^{-4} \, .
\end{equation}
with $+$ for CC neutrino reactions $(n\rightarrow p)$ and $-$ for CC anti-neutrino reactions $(p\rightarrow n)$. Given the smallness of the correction, we shall henceforth set ${\tilde \delta}$ to zero, in which case Eqs.~(\ref{eq-sn-77}) take on exactly the forms used previously in \cite{JandD,Donnelly:2024hqx}. Additionally, from \cite{JandD,Donnelly:2024hqx} we introduce the following:
\begin{eqnarray}
	\alpha_1 &\equiv&  \frac{\sqrt{\tau} {\bar \varepsilon}}{\kappa \sqrt{1 + \tau}}  
	= \sqrt{1 + \frac{\delta^2}{1+\tau}} \nonumber \\
	\alpha_2 &\equiv&  \frac{ \sqrt{\tau} (1 + \tau + {\bar \varepsilon}) }{ 2 \kappa \sqrt{1+\tau} } 
	= \frac{1}{2} \left( \alpha_1 + \frac{\sqrt{\tau (1+\tau)}}{\kappa} \right) . \label{eq-sn-79a} 
\end{eqnarray}
Using results from above we may express the kinematic variables in terms of a single set of three independent quantities; specifically, let us write everything in terms of $(\tau, \eta, \theta)$. We have that $\varepsilon = \sqrt{1+\eta^2}$, $\delta = \eta \sin \theta$ and $\delta' = \eta \cos\theta$, and that the dimensionless 3-momentum transfer and energy transfer may be written
\begin{eqnarray}
	\kappa &=& \frac{\sqrt{\tau}}{1+ \delta^2} \left[ \sqrt{\tau} \delta' + \varepsilon \sqrt{1+\tau+\delta^2} \right] \nonumber \\
	\lambda &=& \frac{\sqrt{\tau}}{1+ \delta^2} \left[ \sqrt{\tau} \varepsilon  + \delta'\sqrt{1+\tau+\delta^2} \right] . \label{eq-nnn1b} 
\end {eqnarray}
See \cite{Donnelly:2024hqx} for other useful identities involving the kinematic variables.

\section{\protect\bigskip Single-Nucleon Axial-vector Current \label{sec-current}}

The discussions in this section all come as extensions to the study presented in \cite{JandD, Donnelly:2024hqx}, where the single-nucleon vector current was the focus. There we saw that 
the single-nucleon on-shell vector current matrix element was given as
\begin{equation}
	J^{\mu }(P^{\prime }\Lambda ^{\prime };P\Lambda )=\bar{u}(P^{\prime
	}\Lambda ^{\prime })\left[ F_1\gamma ^{\mu }+\frac{i}{2 m_N} F_2 \sigma^{\mu\nu}Q_\nu\right] u(P\Lambda ),  \label{sn33v}
\end{equation}
where the initial and final spin projections are denoted $\Lambda$ and $\Lambda'$, respectively, with $\Lambda,\Lambda' = \pm 1/2$, together with expressions for the single-nucleon EM current operator acting between
two-component spin-1/2 spinors, {\it i.e.}, by definition
\begin{equation}
	J^\mu (P\Lambda;P' \Lambda^\prime) \equiv \chi^\dagger_{\Lambda^\prime} {\bar J}^\mu (P;P') \chi_\Lambda . \label{eq-sn-69} 
\end{equation}
The EM form factors are the usual Dirac $(F_1)$ and Pauli $(F_2)$ form factors (see \cite{Musolf:1993tb} for a summary of these quantities and parametrizations for them in other work where the same conventions have been employed, with the exception that in that reference the particle physics convention of having a factor of $\frac{1}{2}$ in defining isoscalar and isovector form factors was adopted rather than the usual nuclear physics convention that is used in the present work, as discussed later). 

Next we briefly summarize results obtained in \cite{JandD} for explicit expressions involving the operator ${\bar J}^\mu$:
\begin{eqnarray}
{\bar J}^\mu &\equiv& f_0 V^\mu \label{eq-sn-80} \\
f_0 &\equiv& \alpha_1 \alpha_2 {\tilde f}_0 \label{eq-sn-81} \\
{\tilde f}_0 &\equiv& \left( \alpha_2^2 +\frac{\tau}{4(1+\tau)} \delta^2 \right)^{-1/2} . \label{eq-sn-82}
\end{eqnarray}
The 4-vector was found to be $V^\mu = (V^0, {\mathbf v})$ with ${\mathbf v} = V^{1^\prime} {\mathbf u}^{\prime}_1 + V^{2^\prime} {\mathbf u}^{\prime}_2 + V^{3^\prime} {\mathbf u}^{\prime}_3$ with
\begin{eqnarray}
V^0 &=& \nu_0 +i \nu'_0 \left( {\bm u}'_2 \cdot {\bm \sigma} \right) \label{eq-cur-1} \\
V^3 &=& \left( \frac{\lambda}{\kappa} \right) V^0 \label{eq-cur-2} \\
{\bm v}^{\perp} &=& \nu_1 {\bm u}'_1 \nonumber \\
&&-i \left\{ \nu'_2 \left( {\bm u}'_3 \times {\bm \sigma} \right) + \nu''_2 \left( {\bm u}'_3 \cdot {\bm \sigma} \right) {\bm u}'_2 + \left( \nu'_2 - \nu'_1 \right) \left( {\bm u}'_2 \cdot {\bm \sigma} \right) {\bm u}'_1 \right\} , \label{eq-cur-3}
\end{eqnarray}
and accordingly 
\begin{eqnarray}
V^0 &=& \nu_0 +i\nu'_0 \sigma^{2'} \label{eq-sn-83} \\
V^{1^\prime} &=& \nu_1 + i \nu'_1 \sigma^{2'} \label{eq-sn-84} \\
V^{2^\prime} &=&  -i \left[ \nu'_2 \sigma^{1'} + \nu''_2 \sigma^{3'} \right]\label{eq-sn-85} \\
V^{3^\prime} &=& \left( \frac{\lambda}{\kappa} \right) V^0 , \label{eq-sn-86} 
\end{eqnarray}
the last arising from the continuity equation. 
The functions $\nu_0$, {\it etc.} are all real and have been derived previously in \cite{JandD}. They are 
\begin{eqnarray}
	{ \nu}_0 &=& \frac{\kappa}{\sqrt{\tau}} \left[ G_E + \frac{\mu_1 \mu_2}{2(1+\tau)}\tau G_M \delta^2 \right] \nonumber \\
	{ \nu}'_0 &=& \frac{\kappa}{\sqrt{1+\tau}} \left[ \mu_1 G_M - \frac{1}{2} \mu_2 G_E \right] \delta \nonumber \\
	{\nu}_1 &=& \frac{1}{\sqrt{1+\tau}} \left[ \mu_1 G_E + \frac{1}{2} \mu_2 \tau G_M \right] \delta\nonumber \\
	{\nu}'_1 &=& \sqrt{\tau} \left[ G_M - \frac{\mu_1 \mu_2}{2(1+\tau)} G_E \delta^2 \right]\nonumber \\
	{\nu}'_2 &=& \sqrt{\tau} \left[ 1 - \frac{\mu_1 \mu_2}{2(1+\tau)} \delta^2 \right] G_M \nonumber \\
	{\nu}''_2 &=& \frac{1}{2} \left( \frac{\lambda}{\kappa} \right) \mu_1 \mu_2 \sqrt{\tau} G_M \delta  \label{eq-sn-92} 
\end{eqnarray}
with $\mu_1 = 1/\alpha_1$ and $\mu_2 = 1/\alpha_2$.

We now adopt the usual assumption of the Conserved Vector Current hypothesis (CVC), taking the vector electromagnetic current operator to be the same as the vector current operator (allowing for the appropriate isospin content) to be employed in weak interaction studies. Note that we could also include a third contribution from second-class scalar currents, but have chosen not to do so in the present work. Furthermore, the form factors may be characterized by their flavor, namely, isoscalar, isovector or strange (see \cite{Musolf:1993tb} for more discussion on this issue).

The axial-vector current may be written in analogy to the vector current as follows:
\begin{equation}
	J^{\mu }_5(P^{\prime }\Lambda ^{\prime };P\Lambda )=\bar{u}(P^{\prime
	}\Lambda ^{\prime })\left[ G_A \gamma ^{\mu }+\frac{1}{2 m_N} G_P Q^\mu\right] \gamma_5 u(P\Lambda ),  \label{defj5}
\end{equation}
where the ``5" is used to indicate axial-vector quantities (from the extra $\gamma_5$ above, compared
with the EM current ) and where $G_A$ and $G_P$ are the axial and induced pseudoscalar form
factors of the nucleon, respectively. (Again, see \cite{Musolf:1993tb} for a summary of these quantities, including discussions of their isospin content and parametrizations for them in other work where the same conventions have been employed).

In the illustrative numerical calculations below, we use the simplest parametrizations for the nucleon form factors. While 
very precise modern parametrizations are available, here we simply aim to demonstrate the significance of different contributions to the
responses. In particular, we investigate the validity of some commonly made approximations involving the quantity $\delta$. Sophisticated form factor
parametrizations are not necessary for this study, and might distract from the point. We use the dipole parametrization
for $G_E^p, G_M^p$ and $G_M^n$
\begin{eqnarray}
	G_E^p (Q^2) = (1 + |Q^2|/M_V^2)^{-2} \nonumber \\
	G_M^p (Q^2) = \mu_p (1 + |Q^2|/M_V^2)^{-2} \nonumber \\
	G_M^n (Q^2) = \mu_n (1 + |Q^2|/M_V^2)^{-2}
\end{eqnarray}
with the magnetic moments $\mu_p = 2.79$ and $\mu_n = -1.91$, and $M_V \approx 0.84$ GeV. For the electric form factor of the neutron, we employ the Galster parametrization
\begin{equation}
	G_E^n (Q^2) = - \frac{\tau \mu_n}{1 + 5.6 \tau} (1 + |Q^2|/M_V^2)^{-2} 
\end{equation}

For the section with numerical results, we present calculations for the charge-changing neutrino reactions, and thus employ the isovector (v) form factors, with 
\begin{equation}
	G_{E,M}^v = G_{E,M}^p - G_{E,M}^n \, ,
\end{equation}  
where we use the nuclear physics convention (with no factor $\frac{1}{2}$ as is sometimes used in particle physics --- see \cite{Musolf:1993tb}) with $G_{M}^v (0) = 4.71$ 
For the axial and induced pseudoscalar form factors, we use the dipole parametrization with the appropriate value $M_A$
\begin{equation}
	G_A (Q^2) = \frac{g_A}{(1 + |Q^2|/M_A^2)^2}
\end{equation}   
with $g_A = 1.267 $. For the induced pseudoscalar form factor $G_P$, we use the assumption of pion pole dominance
parametrization,
\begin{equation}
	G_P (Q^2) = \frac{4 m_N^2}{m_{\pi}^2 + |Q^2|} \,	G_A (Q^2) .
\end{equation}
Note that in the literature, parametrizations of $G_P$ may either show up with a factor of $4 m_N^2$ or $2 m_N^2$. The difference stems
from the way in which the current operator $J^{\mu}_5$ is written down. In our Eq.~(\ref{defj5}), we use $\frac{1}{2 m_N}$ multiplying the
induced pseudoscalar form factor. Other authors choose to write down the current operator with a factor $\frac{1}{m_N}$, leading
to a factor of $2$ instead of $4$ in the parametrization of $G_P$ above. The overall result is obviously the same.

Here we include only first-class currents, omitting second-class axial-vector tensor currents, although they may be included if one desires. For some discussions on second-class scalar V and tensor A currents see, for instance \cite{Walecka,DW,oxygen}. Clearly the pseudoscalar contributions are absent
for transverse projections of the current, since, as usual, the coordinate system employed has
been chosen to have the momentum transfer along the 3-direction. The analog of Eq. (\ref{eq-sn-69})
for the vector current matrix elements becomes the following for the axial-vector current
matrix elements:
\begin{equation}
	J^\mu_5 (P\Lambda;P' \Lambda^\prime) \equiv \chi^\dagger_{\Lambda^\prime} {\bar J}^\mu_5 (P;P') \chi_\Lambda , \label{defj5bar} 
\end{equation}
with
\begin{eqnarray}
	{\bar J}^\mu_5 &\equiv & f_0 A^\mu \nonumber \\
	A^\mu &\equiv& A^\mu_A + A^\mu_P \,,
\end{eqnarray}
where from \cite{JandD,Donnelly:2024hqx} one has
\begin{eqnarray}
	f_0 &\equiv & \alpha_1 \alpha_2 \tilde{f_0}  \nonumber \\
	\tilde{f_0} &\equiv & \left( \alpha_2^2 + \frac{\tau}{4 (1+\tau)} \delta^2 \right )^{-1/2}  \,.
\end{eqnarray}
As in the vector case (see \cite{JandD}) $J^\mu_5$ is a four-vector, whereas $A^\mu$ is not. As stated above,
since the axial-vector current is not conserved, in this case we must treat the $\mu = 0$ and
$\mu = 3$ components of the current independently. One finds from carrying out the procedures
discussed in \cite{JandD,Donnelly:2024hqx} for the EM (vector) current, inserting the spinors and $\gamma$-matrices, that
one has the following results:
\begin{eqnarray}
	A^0 &=& \zeta_0' \bm{\kappa} \cdot \bm{\sigma} + \zeta_0'' \left[  \bm{\eta} - \left ( \frac{\bm{\kappa} \cdot \bm{\eta}}{\kappa^2} \right ) \bm{\kappa} \right] \cdot \bm{\sigma} \nonumber \\
	A^3 &=& \zeta_3' \bm{\kappa} \cdot \bm{\sigma} + \zeta_3'' \left[  \bm{\eta} - \left (\frac{\bm{\kappa} \cdot \bm{\eta}}{\kappa^2} \right ) \bm{\kappa} \right] \cdot \bm{\sigma} \nonumber \\
	\bm{a}^\perp &=& \zeta_1' \bm{\sigma^\perp} + \zeta_2' [(\bm{\kappa} + \bm{\eta}) \cdot \bm{\sigma}] \left[  \bm{\eta} - \left (\frac{\bm{\kappa} \cdot \bm{\eta}}{\kappa^2} \right ) \bm{\kappa} \right] - i \zeta_1 (\bm{\kappa} \times \bm{\eta}) \label{AA}
\end{eqnarray}
with 
\begin{eqnarray}
	\zeta_0' &\equiv& \frac{1}{\alpha_1 \alpha_2} \frac{1}{\sqrt{\tau}} \left ( \frac{\lambda}{\kappa} \right )
	\left \{ G_A' \alpha_3 + \frac{1}{2} G_A \delta^2     \right \}  \nonumber \\
	\zeta_3' &\equiv& \frac{1}{\alpha_1 \alpha_2} \frac{1}{\sqrt{\tau}} 
	\left \{ G_A' \alpha_3 + \frac{1}{2}   \left ( \frac{\lambda}{\kappa} \right )^2    G_A \delta^2     \right \}  \nonumber \\
	\zeta_0'' &\equiv& \frac{1}{\alpha_1 \alpha_2} \frac{\kappa}{\sqrt{\tau}} 
	\left\{ G_A \alpha_3 - \frac{1}{2}   \left ( \frac{\lambda}{\kappa} \right )^2    G_A'     \right \}  \nonumber \\
	\zeta_3'' &\equiv& \frac{1}{\alpha_1 \alpha_2} \frac{\kappa}{\sqrt{\tau}} \left ( \frac{\lambda}{\kappa} \right )
	\left\{ G_A \alpha_3 - \frac{1}{2}    G_A'     \right \}  \nonumber \\
	\zeta_1 &=&  \zeta_2' \equiv \frac{1}{\alpha_1 \alpha_2} \frac{1}{2} \frac{\sqrt{\tau}}{\kappa} G_A \nonumber \\
	\zeta_1' &\equiv& \frac{1}{\alpha_1 \alpha_2} \cdot \sqrt{1 + \tau} G_A \alpha_2	,
\end{eqnarray}
where
\begin{equation}
	\alpha_3 \equiv \alpha_1 \alpha_2 - \frac{1}{2 (1 + \tau) } \delta^2 = \frac{1}{2} \left ( 1 +\frac{\tau}{\kappa^2} \bar{\epsilon} \right ) \,.
\end{equation}
Here we have defined
\begin{equation}
	G_A' \equiv G_A - \tau G_P \label{defgap}
\end{equation}
in analogy with the definition of the electric Sachs form factor $G_E$: as we shall see below,
then only combinations involving $G_A^2$ and $G_A'^2$ enter for certain responses with no cross terms $\sim G_A G_A'$, in the same way
that the introduction of the Sachs form factors eliminated cross-terms proportional to $G_E G_M$
for unpolarized EM scattering. 

\begin{figure}
	\centering
	\includegraphics[width=16cm]{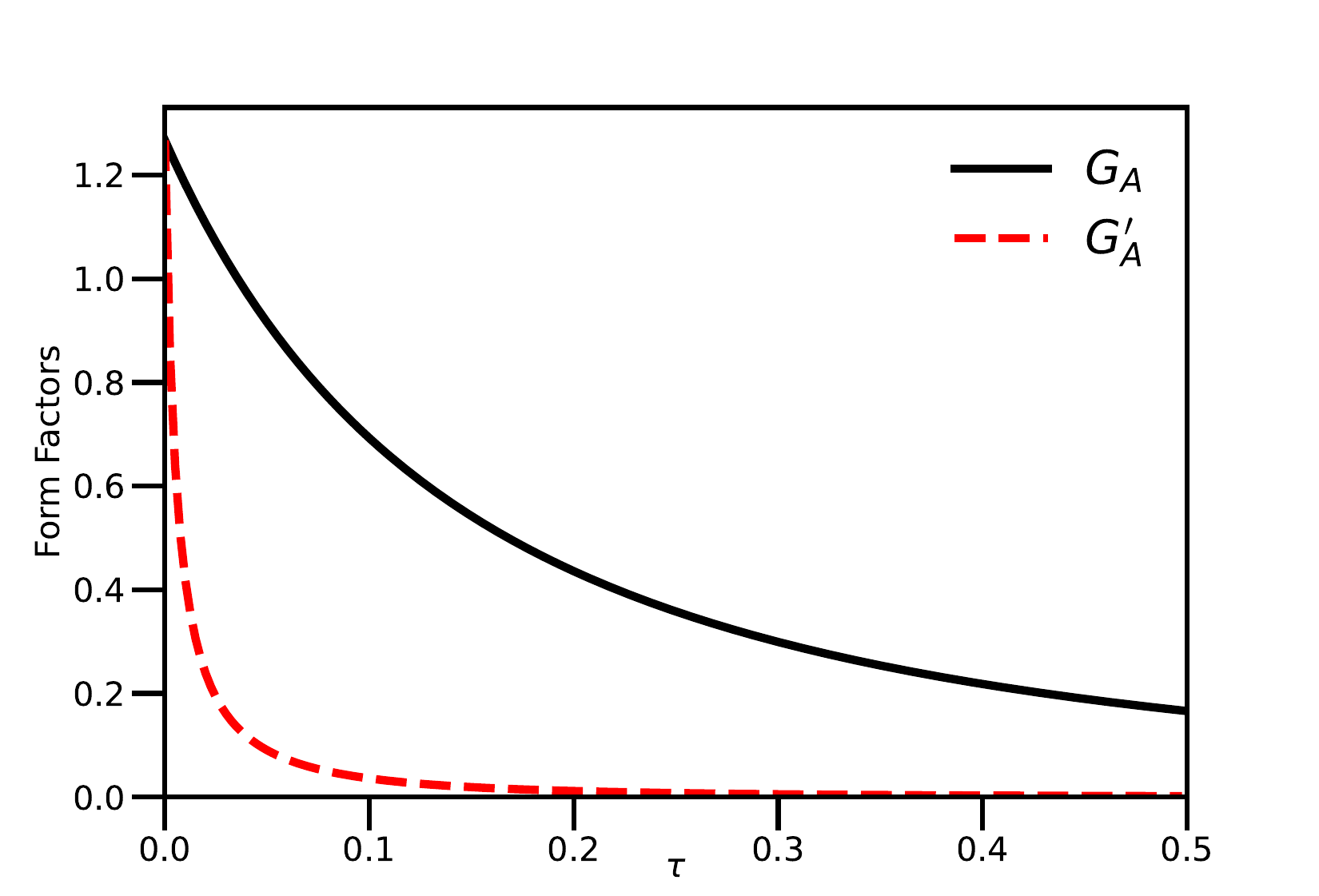} 				
	\caption{The axial form factor $G_A$ (solid line) and the form factor $G_A'$ are shown as a function of $\tau$.  }
	\label{fig_axialffs}
\end{figure}

We use the parametrizations of the axial form factor $G_A$ and the pseudoscalar form factor $G_P$ introduced above. From these, we calculate $G_A'$ as defined above and display the results in Fig. \ref{fig_axialffs}. Looking at the figure, it is obvious that $G_A$ is considerably larger for all values of $\tau$, except for $\tau$ very near zero. Values of $\tau$ at the quasielastic peak, for instance, typically lie above $\tau \sim 0.2$ in studies of neutrino reactions with nuclei, where the single-nucleon current matrix elements are central to modeling the nuclear response functions. This means that we need to consider if a single-nucleon response depends on $G_A$ or $G_A'$, as this introduces a very significant difference in size; as will be discussed in Sect.~\ref{sec-results}, this leads to an excellent approximation for typical kinematics, namely, to ignore any contributions that involve $G_A'$. 

Note also that in Eqs.~(\ref{AA}) we must treat the $\mu = 0$ and $\mu = 3$ components
separately, since the axial-vector current is not conserved, in contrast to the vector case
where the continuity equation relates one to the other. 

As in the vector case (see \cite{JandD,Donnelly:2024hqx}) the spin dependence may be made explicit.  Note that we are using the single-primed coordinate system discussed in \cite{Donnelly:2024hqx} and used in \cite{JandD}, with $\bm{u}_3' \equiv \bm{\kappa}/\kappa$, 
	$\bm{u}_2' \equiv (\bm{\kappa} \times \bm{\eta})/ \kappa \eta \sin \theta $ and $\bm{u}_1' \equiv \bm{u}_2' \times \bm{u}_3'$:
\begin{eqnarray}
	A^0 &=& \beta_1' \sigma^{1'} + \beta_3' \sigma^{3'} \nonumber \\
	A^3 &=& \beta_1'' \sigma^{1'} + \beta_3'' \sigma^{3'} \nonumber \\
	\bm{a}^\perp &=& \left [ (\gamma_1' \sigma^{1'} + \gamma_3' \sigma^{3'} ) \bm{u}_1' + \gamma_2' \sigma^{2'} \bm{u}_2' \right]  -i \gamma_2 \bm{u}_2' ,
\end{eqnarray}
with
\begin{eqnarray}
	\beta_1' &\equiv& \delta \zeta_0'' \nonumber \\
	\beta_3' &\equiv& \kappa \zeta_0' \nonumber \\
	\beta_1'' &\equiv& \delta \zeta_3'' \nonumber \\
	\beta_3'' &\equiv& \kappa \zeta_3' \nonumber \\
	\gamma_1' &\equiv& \frac{1}{\alpha_2} \sqrt{1 + \tau} G_A \alpha_3 \nonumber \\
	\gamma_3' &\equiv& \frac{1}{2 \alpha_2} \left ( \frac{\lambda}{\kappa} \right )  \sqrt{1 + \tau} G_A \delta \nonumber \\
	\gamma_2' &\equiv& \zeta_1' = \frac{1}{\alpha_1} \cdot \sqrt{1 + \tau} G_A \nonumber \\
	\gamma_2 &\equiv& \kappa \delta \zeta_1 = \frac{1}{2 \alpha_1 \alpha_2} \sqrt{\tau} G_A \delta .
	\label{defbetagammas}
\end{eqnarray}
The above allows us to rewrite the expressions for the components of $A^{\mu}$ in the primed coordinate system, 
and to sort the resulting terms according to the power of $\delta$ they contain.

\begin{eqnarray}
	A^0 &=& \frac{1}{\alpha_1 \alpha_2} \, \frac{\kappa}{\sqrt{\tau}} \, \alpha_3
	 \left \{ \frac{\lambda}{\kappa}  G_A' (\bm{u}_3' \cdot \bm{\sigma}) + \delta \left [ G_A -  \left( \frac{\lambda}{\kappa} \right)^2
	 \frac{1}{2 \alpha_3} G_A' \right ] (\bm{u}_1' \cdot\bm{\sigma}) 
	 + \delta^2 \frac{\lambda}{\kappa} \frac{1}{2 \alpha_3} G_A (\bm{u}_3' \cdot \bm{\sigma})  \right \} \nonumber \\
	A^{1'} &=& \frac{\alpha_3}{\alpha_2} \sqrt{1 + \tau} G_A \left \{ (\bm{u}_1' \cdot\bm{\sigma}) 
	+ \delta \, \frac{\lambda}{\kappa} \, \frac{1}{2 \alpha_3} (\bm{u}_3' \cdot \bm{\sigma})  \right \}	\nonumber \\ 
	A^{2'} &=& \frac{1}{\alpha_1} G_A \left \{ \sqrt{1 + \tau} (\bm{u}_2' \cdot \bm{\sigma})  - \delta i \, \frac{1}{2 \alpha_2} \sqrt{\tau}  \right \} \nonumber \\
	 A^{3'} &=& \frac{1}{\alpha_1 \alpha_2} \, \frac{\kappa}{\sqrt{\tau}}  \, \alpha_3
	 \left \{ G_A' (\bm{u}_3' \cdot \bm{\sigma}) + \delta \frac{\lambda}{\kappa}  \left [ G_A - \frac{1}{2 \alpha_3} G_A'  \right ]
	 (\bm{u}_1' \cdot\bm{\sigma}) + \delta^2 \left( \frac{\lambda}{\kappa} \right)^2 \frac{1}{2 \alpha_3} G_A (\bm{u}_3' \cdot \bm{\sigma}) \right \} \nonumber \\
\end{eqnarray}
While the terms in the above equations are sorted according to the power of $\delta$, these equations are fully correct, covariant expressions without any approximations. They are not expansions and are correct at all energy scales. We consider the leading terms in $\delta$ in the next subsection.

\subsection{Expansion in  $\eta$}

The results in Eqs. (\ref{defbetagammas}) can be expanded in powers of $\eta$:
\begin{eqnarray}
	\beta_1' &=& \frac{\kappa}{\sqrt{\tau}} \left [ G_A - \frac{1}{2} \left ( \frac{\lambda}{\kappa} \right )^2 G_A' \right ]
	\eta \sin \theta + {\cal O}(\eta^2) \nonumber \\
	\beta_1'' & =& \frac{\lambda}{\sqrt{\tau}} \left [ G_A -\frac{1}{2} G_A' \right ] \eta \sin \theta + {\cal O}(\eta^2) \nonumber \\
	\beta_3' &=& \frac{\lambda}{\sqrt{\tau}} G_A' + {\cal O}(\eta^2) \nonumber \\
	\beta_3'' & =& \frac{\kappa}{\sqrt{\tau}} G_A' + {\cal O}(\eta^2) \nonumber \\
	\gamma_2 &=& \frac{1}{2} \sqrt{\tau} G_A \eta \sin \theta + {\cal O}(\eta^2) \nonumber \\
	\gamma_1' &=& \gamma_2' = \sqrt{1 + \tau} G_A + {\cal O}(\eta^2) \nonumber \\
	\gamma_3' &=& \frac{1}{2} \sqrt{\tau} G_A \eta \sin \theta + {\cal O}(\eta^2) .	
\end{eqnarray}
The result is that the axial-vector current operator to order $\eta$ is given by
\begin{eqnarray}
	\bar{J}_5^0 &=& \frac{\lambda}{\sqrt{\tau}} G_A' \left (  \frac{1}{\kappa} \bm{\kappa} \cdot \bm{\sigma}\right ) \nonumber \\
	  & & +\sqrt{1+\tau} \left ( G_A -\frac{1}{2} \left (\frac{\tau}{1+\tau} \right) G_A' \right ) 
	  \left[  \bm{\eta} - \left (\frac{\bm{\kappa} \cdot \bm{\eta}}{\kappa^2} \right ) \bm{\kappa} \right] \cdot \bm{\sigma} \nonumber \\
	  \bar{J}_5^3 &=& \frac{\kappa}{\sqrt{\tau}} G_A' \left (  \frac{1}{\kappa} \bm{\kappa} \cdot \bm{\sigma}\right ) \nonumber \\
	  & & +\sqrt{\tau} \left( G_A - \frac{1}{2} G_A' \right) \left[  \bm{\eta} - \left (\frac{\bm{\kappa} \cdot \bm{\eta}}{\kappa^2} \right ) \bm{\kappa} \right] \cdot \bm{\sigma} \nonumber \\
	  \bar{J}_5^\perp &=& \sqrt{1+\tau} G_A \bm{\sigma^\perp} \nonumber \\
	   & & +\frac{1}{2 \sqrt{1 + \tau}} G_A \left ( (\bm{\kappa} \cdot \bm{\sigma})  \left[  \bm{\eta} - \left (\frac{\bm{\kappa} \cdot \bm{\eta}}{\kappa^2} \right ) \bm{\kappa} \right]  - i (\bm{\kappa} \times \bm{\sigma}) \right ) ,
\end{eqnarray}
where one has
\begin{eqnarray}
	\lambda &=& \tau + \sqrt{\tau (1+\tau)} \delta' + {\cal O}(\eta^2) \nonumber \\
	\kappa &=& \sqrt{\tau (1+\tau)} + \tau  \delta ' + {\cal O}(\eta^2)
	\label{lamkapfirstorder}
\end{eqnarray}
and thus the contributions that occur in leading order may be expanded to linear order
using
\begin{eqnarray}
	\frac{\lambda}{\sqrt{\tau}} &=& \sqrt{\tau} + \sqrt{1+\tau} \delta' + {\cal O}(\eta^2) \nonumber \\
	\frac{\kappa}{\sqrt{\tau}} &=& \sqrt{1+\tau} + \sqrt{\tau} \delta' + {\cal O}(\eta^2) .
	\label{kinfacfirstorder}
\end{eqnarray}
For the contributions that are explicitly ${\cal O}(\eta)$ in the current we have evaluated the kinematic factors with $\eta$ set to zero, leading to the coefficients $\sqrt{1+\tau}$ and $\sqrt{\tau}$ in the second terms above containing $\delta'$. 

As discussed in more detail in \cite{Donnelly:2024hqx} the usual prescription for nuclear physics is to replace ${\mathbf \eta}$ with $-i{\mathbf \nabla}/m_N$ and work in coordinate space. Typically only the leading order and some, but not all, corrections ${\cal O}(\eta)$ are included, as pointed out in that reference, which can have consequences when $q$ and $\omega$ are large, as is often the case in quasielastic scattering.

\section{Electroweak Single-nucleon response functions\label{sec-responses}}

We now proceed to a presentation of the ingredients needed in descriptions of charge-changing neutrino and anti-neutrino reactions, where therefore only isovector form factors are required. In \cite{Moreno:2014kia} the general form for semi-inclusive charged-current neutrino-nucleus reactions was presented. We can draw on those developments in our present study, since some of those past results were independent of the specific nature of the reaction. In particular, the cross section is found to be proportional to the contraction of the leptonic and hadronic tensors involved, namely to a general form given by
\begin{equation}
v_0 \:\mathcal{F}_{\chi}^2 \equiv \eta _{\mu \nu }W^{\mu \nu } = \eta _{\mu \nu }^{s}W_{s}^{\mu \nu } + \chi \:\eta _{\mu \nu }^{a}W_{a}^{\mu \nu } \:,
\end{equation}
where $\chi = 1$ for incident neutrinos and $\chi = -1$ for anti-neutrinos. The symmetric and  antisymmetric components of the leptonic and the hadronic tensors can be contracted separately, since no cross-terms are allowed.  

The leptonic tensors $\eta_{\mu\nu}^{s,a}$ are given in \cite{Moreno:2014kia} for the general situation in which mass terms are retained. Note that in that reference the symbols $\delta$ and $\delta'$ were defined differently from how they are used in the present study and in previous work upon which this work is based. The present case of CC neutrino or anti-neutrino reactions means that the leptonic coupling constants are $a_V = +1$ and $a_A = -1$. In passing we note that, were one to want expressions here for neutral current neutrino or anti-neutrino reactions, then $a_{V,A}$ would be different and involve the weak mixing angle, as discussed in other work. 
We employ the familiar notation in which the contractions may be written as sums of contributions projected onto charge (C), longitudinal (L) and transverse (T) terms, namely, $\widehat{V}_{CC}$, {\it etc.} (see also \cite{Moreno:2014kia} for explicit forms for these quantities). 
The hadronic tensors may likewise be similarly decomposed into response functions, namely, $W^{CC}$, {\it etc.}; for the case of single-nucleon reactions these provide the focus for the discussions below in the present section. The relevant contractions are then the following:
\begin{eqnarray}
\eta _{\mu \nu }^{s}W_{s}^{\mu \nu } &=& v_{0}\left\{ \widehat{V}_{CC}W^{CC}+\widehat{V}_{CL}W^{CL}+\widehat{V}_{LL}W^{LL}\right.  \nonumber \\
&& \left. +\widehat{V}_{T}W^{T}+\widehat{V}_{TT}W^{TT}  +\widehat{V}_{TC}W^{TC}+\widehat{V}_{TL}W^{TL} \right\}  \nonumber
\end{eqnarray}
\begin{eqnarray}
\eta _{\mu \nu }^{a}W_{a}^{\mu \nu } &=& v_{0}\left\{ \widehat{V}_{T^{\prime }}W^{T^{\prime }}+\widehat{V}_{TC^{\prime }}W^{TC^{\prime }}+\widehat{V}_{TL^{\prime }}W^{TL^{\prime }} 
\right\} ,  \label{eqcon31}
\end{eqnarray}
where the overall factor $v_0$ is given in \cite{Moreno:2014kia}.

These hadronic tensors may be written with explicit contributions from V and A current matrix elements, namely, for the symmetric tensors
\begin{equation}
W^i = W^i_{VV} + W^i_{AA}
\end{equation}
with $i= CC, CL, LL, T, TT, TC$ and $TL$, and for the anti-symmetric tensors
\begin{equation}
W^i = W^i_{VA} + W^i_{AV} = 2W^i_{VA}
\end{equation}
with $i= T', TC'$ and $TL'$, the factor of 2 coming from the symmetry in the contraction of the tensors. This leads to the following sets of contractions:
\begin{eqnarray}
\eta _{\mu \nu }^{s}\left[ W_{s}^{\mu \nu }\right]_{VV} &=& v_{0}\left\{ \widehat{V}_{CC}W^{CC}_{VV}+\widehat{V}_{CL}W^{CL}_{VV}+\widehat{V}_{LL}W^{LL}_{VV}\right.  \nonumber \\
&& \left. +\widehat{V}_{T}W^{T}_{VV}+\widehat{V}_{TT}W^{TT}_{VV}  +\widehat{V}_{TC}W^{TC}_{VV}+\widehat{V}_{TL}W^{TL}_{VV} \right\} \label{sVV}
\end{eqnarray}
\begin{eqnarray}
\eta _{\mu \nu }^{s}\left[ W_{s}^{\mu \nu }\right]_{AA} &=& v_{0}\left\{ \widehat{V}_{CC}W^{CC}_{AA}+\widehat{V}_{CL}W^{CL}_{AA}+\widehat{V}_{LL}W^{LL}_{AA}\right.  \nonumber \\
&& \left. +\widehat{V}_{T}W^{T}_{AA}+\widehat{V}_{TT}W^{TT}_{AA}  +\widehat{V}_{TC}W^{TC}_{AA}+\widehat{V}_{TL}W^{TL}_{AA} \right\} \label{sAA}
\end{eqnarray}
\begin{eqnarray}
\eta _{\mu \nu }^{a}\left[ W_{a}^{\mu \nu } \right]_{VA+AV}&=& 2v_{0}\left\{ \widehat{V}_{T^{\prime }}W^{T^{\prime }}_{VA}+\widehat{V}_{TC^{\prime }}W^{TC^{\prime }}_{VA}+\widehat{V}_{TL^{\prime }}W^{TL^{\prime }}_{VA} 
\right\} .  \label{eqcon31a}
\end{eqnarray}

We begin the development of the hadronic single-nucleon response functions by recalling the results for the purely vector case
already obtained in previous work \cite{JandD,Donnelly:2024hqx}; here these are denoted ``VV" to distinguish them
from the other contributions discussed later. We consider only the spin-saturated (unpolarized)
responses (see \cite{Donnelly:2024hqx}), and in the present work we have not detailed the relationships
between the 123-system and the $1'2'3'$-system as these may be found in \cite{Donnelly:2024hqx}. Upon implementing
these relationships we find the factors involving the angle $\phi$, as was found in the purely
vector case. Since the hadronic vector current is assumed to be conserved we have that
\begin{eqnarray}
W^{CL}_{VV} &=& 2 Re W^{03}_{VV} = 2 \left( \frac{\lambda}{\kappa} \right ) W^{CC}_{VV} \nonumber \\
W^{LL}_{VV} &=& Re W^{33}_{VV} = \left( \frac{\lambda}{\kappa} \right )^2 W^{CC}_{VV} \nonumber \\
W^{TL}_{VV} &=& 2 \sqrt{2} Re W^{31}_{VV} = \left( \frac{\lambda}{\kappa} \right ) W^{TC}_{VV} = \left( \frac{\lambda}{\kappa} \right ) 2 \sqrt{2} Re W^{01}_{VV} .\label{eq-CC}
\end{eqnarray}
This allows us to define
\begin{eqnarray}
\widehat{V}_{(L)} &\equiv& \widehat{V}_{CC} + 2 \left( \frac{\lambda}{\kappa} \right ) \widehat{V}_{CL} + \left( \frac{\lambda}{\kappa} \right )^2 \widehat{V}_{LL} \nonumber \\
\widehat{V}_{(TL)} &\equiv& \widehat{V}_{TC} + \left( \frac{\lambda}{\kappa} \right ) \widehat{V}_{TL} , \label{eq-Vs}
\end{eqnarray}
and then in the VV sector we have only the combinations $\widehat{V}_{(L)} W^{CC}_{VV}$ and $\widehat{V}_{(TL)} W^{TC}_{VV}$, together with
\begin{eqnarray}
	  W^{22}_{VV} + W^{11}_{VV} &\equiv& W^T_{VV} \nonumber \\
	  W^{22}_{VV} - W^{11}_{VV} &\equiv& W^{TT}_{VV} 
\end{eqnarray}
multiplied by $\widehat{V}_{T}$ and $\widehat{V}_{TT}$, respectively.
From \cite{JandD,Donnelly:2024hqx} we have
\begin{eqnarray}
	W^{CC}_{VV} &=& \frac{\kappa^2}{\tau} \left[ G_E^2 + \delta^2 W_2 \right]  \nonumber \\
	W^T_{VV} &=& 2W_1 + \delta^2 W_2  \nonumber \\
	W^{TT}_{VV} &=&  -\delta^2 W_2 \cos 2\phi  \nonumber \\
	W^{TC}_{VV} &=& 2\sqrt{2} {\bar \varepsilon} \delta W_2  \cos \phi  \label{vvresps}
\end{eqnarray}
with $W_1 \equiv \tau G_M^2$ and $W_2 \equiv [G_E^2 + \tau G_M^2]/(1+\tau) $. 

The spin-saturated products of matrix elements of the axial-vector current operator expressions
(denoted ``AA") are obtained in a similar way: one obtains the following:
\begin{eqnarray}
	W^{CC}_{AA} &\equiv& W_{AA}^{00} = f_0^2 (\beta_1'^2 + \beta_3'^2) = \frac{\kappa^2}{\tau} 
	\left \{ \left (\frac{\lambda}{\kappa} \right)^2 G_A'^2 + G_A^2 \delta^2  \right \} \nonumber \\
	W^{LL}_{AA} &\equiv& W_{AA}^{33} = f_0^2 (\beta_1''^2 + \beta_3''^2) = \frac{\kappa^2}{\tau} 
	\left \{  G_A'^2 + \left (\frac{\lambda}{\kappa} \right)^2 G_A^2 \delta^2  \right \} \nonumber \\
	W^{CL}_{AA} &\equiv& 2W_{AA}^{03} = 2f_0^2 (\beta_1' \beta_1'' + \beta_3' \beta_3'') = 2\frac{\kappa^2}{\tau} 
	 \left (\frac{\lambda}{\kappa} \right) \left \{  G_A'^2 + G_A^2 \delta^2  \right \} \nonumber \\
	 W^T_{AA} & = & f_0^2 [(\gamma_2^2 + \gamma_2'^2) +(\gamma_1'^2 + \gamma_3'^2)] = [2 (1+\tau) + \delta^2] G_A^2 \nonumber \\
	 W^{TT}_{AA} & = & f_0^2 [(\gamma_2^2 + \gamma_2'^2) - (\gamma_1'^2 + \gamma_3'^2)]  \cos 2 \phi =
	 - \delta^2 G_A^2 \cos 2 \phi \nonumber \\
	 W^{TC}_{AA} &\equiv&  2 \sqrt{2} W^{01}_{AA} = 2 \sqrt{2} f_0^2 (\beta_1' \gamma_1' + \beta_3' \gamma_3') \cos \phi = 2 \sqrt{2} \bar{\epsilon} G_A^2 
	 \delta \cos \phi \nonumber \\
	 W^{TL}_{AA} &\equiv& 2 \sqrt{2} W^{31}_{AA} = 2 \sqrt{2} f_0^2 (\beta_1'' \gamma_1' + \beta_3'' \gamma_3') \cos \phi = 2 \sqrt{2}
	 \left (\frac{\lambda}{\kappa} \right) \bar{\epsilon} G_A^2 
	 \delta \cos \phi . \label{aaresps}
\end{eqnarray}
Note that only the combinations $G_A^2$ and $G_A'^2$
 occur here with no interference terms proportional
to $G_A G_A'$; a similar situation was found in discussing the purely vector current
operators (see above), where in that case only contributions proportional to $G_E^2$ and $G_M^2$
enter with no interference terms proportional to $G_E G_M$ for such spin-saturated results. In
this sense the choice of $G_A'$ made above is analogous to the definition of the Sachs form
factors.
Finally, when the product of vector and axial-vector contributions is considered (denoted
``VA") one has the following:
\begin{eqnarray}
	W^{T'}_{VA} &\equiv & - 2 {\rm Im} (W^{12}_{VA}) = 2 f_0^2 (\nu_1 \gamma_2 + \nu_1' \gamma_2') \nonumber \\
	&=& 2 {\rm Im} (W^{21}_{VA}) = 2 f_0^2 (\nu_2' \gamma_1' + \nu_2'' \gamma_3') \nonumber \\
	&=& 2 \frac{\tau}{\kappa} \bar{\epsilon} G_M G_A \nonumber \\
	W^{TC'}_{VA} &\equiv& - 2 \sqrt{2} {\rm Im} (W^{02}_{VA}) \sin \phi = 2 \sqrt{2} f_0^2 (\nu_0 \gamma_2 + \nu_0' \gamma_2') 
	\sin \phi \nonumber \\
	&=& 2 \sqrt{2} {\rm Im} (W^{20}_{VA}) \sin \phi = 2 \sqrt{2} f_0^2 (\nu_2' \beta_1' + 
	\nu_2'' \beta_3'') 
	\sin \phi \nonumber \\
	&=& 2 \sqrt{2} \kappa G_M G_A \delta \sin \phi	 \nonumber \\
	W^{TL'}_{VA} &\equiv& - 2 \sqrt{2} {\rm Im} (W^{32}_{VA}) \sin \phi = \left ( \frac{\lambda}{\kappa} \right ) 
	W^{TC'}_{VA} \nonumber \\
	&=& 2 \sqrt{2} {\rm Im} (W^{23}_{VA}) \sin \phi = 2 \sqrt{2} f_0^2 (\nu_2' \beta_1'' + \nu_2'' \beta_3'')
	\sin \phi \nonumber \\
	&=& 2 \sqrt{2} \lambda G_M G_A \delta \sin \phi  \label{varesps} \, .
\end{eqnarray}
The other terms, $W^{T'}_{AV}$, $W^{TC'}_{AV}$ and $W^{TL'}_{AV}$ are easily obtained either directly or using the symmetry in the tensors. Here only the combination $G_M G_A$ occurs, with no terms proportional to $G_M G_A'$, $G_E G_A$ or
$G_E G_A'$.
All other responses are zero (see \cite{Donnelly:2023rej,arxivlong} for some discussions of responses that are even under time-reversal (TRE) and odd under time-reversal (TRO) responses; here only TRE responses are non-zero).

Again we emphasize the fact that these are fully correct, covariant expressions for single-nucleon neutrino or anti-neutrino reactions at all energy scales where the struck nucleon is moving. No expansions have been made; those are presented in the following subsection.

\subsection{Expansions in $\eta$ \label{sec-respsexpand}}

The general results given above can all be expanded in powers of $\eta$; here we expand to
linear order, dropping all contributions of $ {\cal O}(\eta^2)$  and higher: for the VV contributions we
have (see \cite{JandD,Donnelly:2024hqx})
\begin{eqnarray}
	W^L_{VV} &=& \frac{\kappa^2}{\tau} G_E^2 + {\cal O}(\eta^2) \nonumber \\
	W^T_{VV} &=& 2W_1 +  {\cal O}(\eta^2)\nonumber \\
	W^{TT}_{VV} &=&   {\cal O}(\eta^2) \nonumber \\
	W^{TL}_{VV} &=& 2\sqrt{2} (1+\tau) \delta W_2  \cos \phi + {\cal O}(\eta^2) \,.
\end{eqnarray}
Note that in the last equation above we have used the fact that it is already of order $\eta$ to
simplify the result. We may also expand the factor $\kappa^2/\tau$ in powers of $\eta$; see the discussions
in the detailed results to follow.

Similarly for the AA contributions we have
\begin{eqnarray}
	W^{CC}_{AA} &=& \frac{\lambda^2}{\tau} G_A'^2  + {\cal O}(\eta^2) \nonumber \\
	W^{LL}_{AA} &=& \frac{\kappa^2}{\tau} G_A'^2  + {\cal O}(\eta^2) \nonumber \\
	W^{CL}_{AA} &=& 2 \frac{\lambda \kappa}{\tau} G_A'^2  + {\cal O}(\eta^2) \nonumber \\
	W^T_{AA} &=& 2 (1+\tau) G_A^2 + {\cal O}(\eta^2) \nonumber \\
	W^{TT}_{AA} &=& {\cal O}(\eta^2) \nonumber \\
	W^{TC}_{AA} &=& 2 \sqrt{2} (1+\tau) G_A^2 \delta \cos \phi + {\cal O}(\eta^2) \nonumber \\
	W^{TL}_{AA} &=& 2 \sqrt{2} \sqrt{\tau (1+\tau)} G_A^2 \delta \cos \phi + {\cal O}(\eta^2) ,
\end{eqnarray}
where, in addition to the explicitly ${\cal O}(\eta)$ contributions that arise because $\delta = \eta \sin \theta$, one may also expand the factors in the first three contributions using the expansions in Eqs. (\ref{kinfacfirstorder}).

For the VA contributions we have
\begin{eqnarray}
	W^{T'}_{VA} &=& 2 \sqrt{\tau (1+\tau)} G_M G_A + {\cal O}(\eta^2) \nonumber \\
	W^{TC'}_{VA} &=& 2 \sqrt{2} \sqrt{\tau (1+\tau)} G_M G_A \delta \sin \phi + {\cal O}(\eta^2) \nonumber \\
	W^{TL'}_{VA} &=& 2 \sqrt{2} \tau G_M G_A \delta \sin \phi + {\cal O}(\eta^2) \,.
\end{eqnarray}
Finally, by setting $\eta$ to zero we can obtain the rest-frame (R) results:
\begin{eqnarray}
	\left [W^L_{VV} \right]_R &=& (1+\tau) G_E^2 \nonumber \\
	\left [ W^T_{VV} \right]_R  &=&  2 W_1 \nonumber \\
	\left [W^{CC}_{AA} \right]_R &=& \tau G_A'^2 \nonumber \\
	\left [W^{LL}_{AA} \right]_R &=& (1 +\tau) G_A'^2 \nonumber \\
	\left [W^{CL}_{AA} \right]_R &=& 2\sqrt{\tau (1+\tau)} G_A'^2 \label{eq-zeros} \\
	\left [W^T_{AA} \right]_R &=& 2 (1+\tau) G_A^2 \nonumber \\
	\left [W^{T'}_{VA} \right]_R &=& 2 \sqrt{\tau (1+\tau)} G_M G_A \nonumber \\
	\left [W^{TT}_{VV} \right]_R &=& \left[W^{TL}_{VV}\right]_R = \left[W^{TT}_{AA}\right]_R = \left[W^{TC}_{AA}\right]_R = \left[W^{TL}_{AA}\right]_R = \left[W^{TC'}_{VA} \right]_R = \left[W^{TL'}_{VA}\right]_R = 0 \,. \nonumber
\end{eqnarray}

\subsection{Breit Frame Responses \label{sec-respbreit}}

We finish this section with the results for collinear kinematics. 
As is well-known in studying elastic electron scattering including from nucleons as targets, it proves useful to cast the results in the so-called Breit frame (``B''). In this frame one has the target 3-momentum {\bf p}$_B$ collinear with the 3-momentum transfer {\bf q} and opposed to it, and has the final-state 3-momentum {\bf p}$^\prime_B$ in the direction of {\bf q} and equal in magnitude to that of {\bf p}$_B$. This implies that the initial and final 3-momenta are $p_B=p'_B=q_B/2$ and therefore that the corresponding nucleon energies are $E_{p,B}=E'_{p,B}=\sqrt{m_N^2+q_B^2/4}$. For the convenience of the reader, we list the kinematics variables evaluated in the Breit frame: 
\begin{eqnarray}
	{\bm \eta}_B &=& - {\bm \kappa}_B \nonumber \\ 
	\theta_B &=& \pi \nonumber \\ 
	\delta_B &=& 0 \nonumber \\ 
	\delta'_B &=& - \eta_B \nonumber \\ 
	\omega_B &=& 0 \nonumber \\ 
	\lambda_B &=& 0 \nonumber \\ 
	\kappa_B &=& \sqrt{\tau} \qquad\qquad \kappa_B/\sqrt{\tau} = 1 \nonumber \\ 
	\varepsilon_B &=& \sqrt{1 + \kappa_B^2} = \sqrt{1 + \tau} , \label{breitkinematics}
\end{eqnarray}
where $\theta_B = \pi$ 
implies that $\sin \theta_B = 0$ and $\cos \theta_B = -1$.  This yields the following for the 4-vectors employed in this study:
\begin{eqnarray}
		Q^\mu_B &=& q_B (0,0,0,1) = 2 m_N \sqrt{\tau} (0,0,0,1) \nonumber \\ 
		P^\mu_B &=& (\sqrt{m_N^2+q_B^2/4},0,0,-q_B/2) = m_N (\sqrt{1+\tau},0,0,-\sqrt{\tau} ) \label{eq-B-10}  .  \label{eq-B-11} 
\end{eqnarray}
		
These results allow one to write the required responses in the Breit frame for the unpolarized cases using Eqs.~(\ref{vvresps}) together with the identities $\kappa_B = \sqrt{\tau}$, $\delta_B=0$ from above. From \cite{Donnelly:2024hqx}, we have for the VV responses
\begin{eqnarray}
			\left[ W^L_{VV} \right]_B &=& G_E^2 \nonumber \\ 
			\left[ W^T_{VV} \right]_B &=& 2 \tau G_M^2 \nonumber \\ 
			\left[ W^{TT}_{VV} \right]_B &=& \left[ W^{TL}_{VV} \right]_B =0 , \label{eq-B-29} 
\end{eqnarray}
and the AA responses, using Eqs.~(\ref{aaresps}),  in the Breit frame are
\begin{eqnarray}
	 \left[ W^{LL}_{AA} \right]_B &=& G_A'^2 \nonumber \\ 
	 \left[ W^{T}_{AA} \right]_B &=& 2 (1 + \tau) G_A^2 \nonumber \\ 
	 \left[ W^{CC}_{AA} \right]_B &=& \left[ W^{CL}_{AA} \right]_B = \left[ W^{TT}_{AA} \right]_B = \left[ W^{TC}_{AA} \right]_B
	 = \left[ W^{TL}_{AA} \right]_B = 0 \,.
\end{eqnarray}
For the VA responses, Eqs.~(\ref{varesps}), only one is non-zero in the Breit frame:
\begin{eqnarray}
	\left[ W^{T'}_{VA} \right]_B &=& 2 \sqrt{\tau (1+\tau)} G_M G_A \nonumber \\
	\left[ W^{TC'}_{VA} \right]_B &=& \left[ W^{TL'}_{VA} \right]_B = 0 \,.
\end{eqnarray}

\section{Detailed Results for single-nucleon response functions\label{sec-results}}

All results in this section are chosen for use in CC neutrino and anti-neutrino reactions, namely, purely isovector responses, although it is straightforward to extend the ideas presented here to other electroweak reactions, for instance, to NC neutrino or anti-neutrino scattering. We present numerical results for a few selected kinematical choices to illustrate the typical scales of the response functions for several choices of $\tau$, $\theta$ and $\phi$ as functions of $\eta$. We draw on the insights gotten in \cite{Donnelly:2024hqx} where we argued on the basis of the Relativistic Fermi Gas (RFG) model \cite{Alberico:1988bv} to specify the ranges of kinematics in which one should reasonably expect nuclear cross sections to be significant. Specifically, if the RFG scaling variable is restricted to the range $-0.7 < \psi < +0.7$ for the RFG scaling variable, then one covers the region around the quasielastic peak (which occurs near $\psi = 0$) where the nuclear cross sections are large. From \cite{Donnelly:2024hqx} and below in Fig. \ref{fig-kinematicsall} we include figures to indicate the ranges one finds for the entire set of kinematic variables, including $\psi$, as functions of $\eta$. 

We compare the ``full results'' with two specific first-order approximations: one is called the ``single first-order approximation'' wherein factors involving the ratio $\kappa/\sqrt{\tau}$ are not approximated, but where higher-order contributions in $\eta$ to the responses are neglected, and a second called the ``double first-order approximation'' in which the factors involving $\kappa/\sqrt{\tau}$ are also expanded to linear order in $\eta$, see Eqs.~(\ref{lamkapfirstorder},\ref{kinfacfirstorder}).

\begin{figure}
	\centering
	\includegraphics[width=18cm]{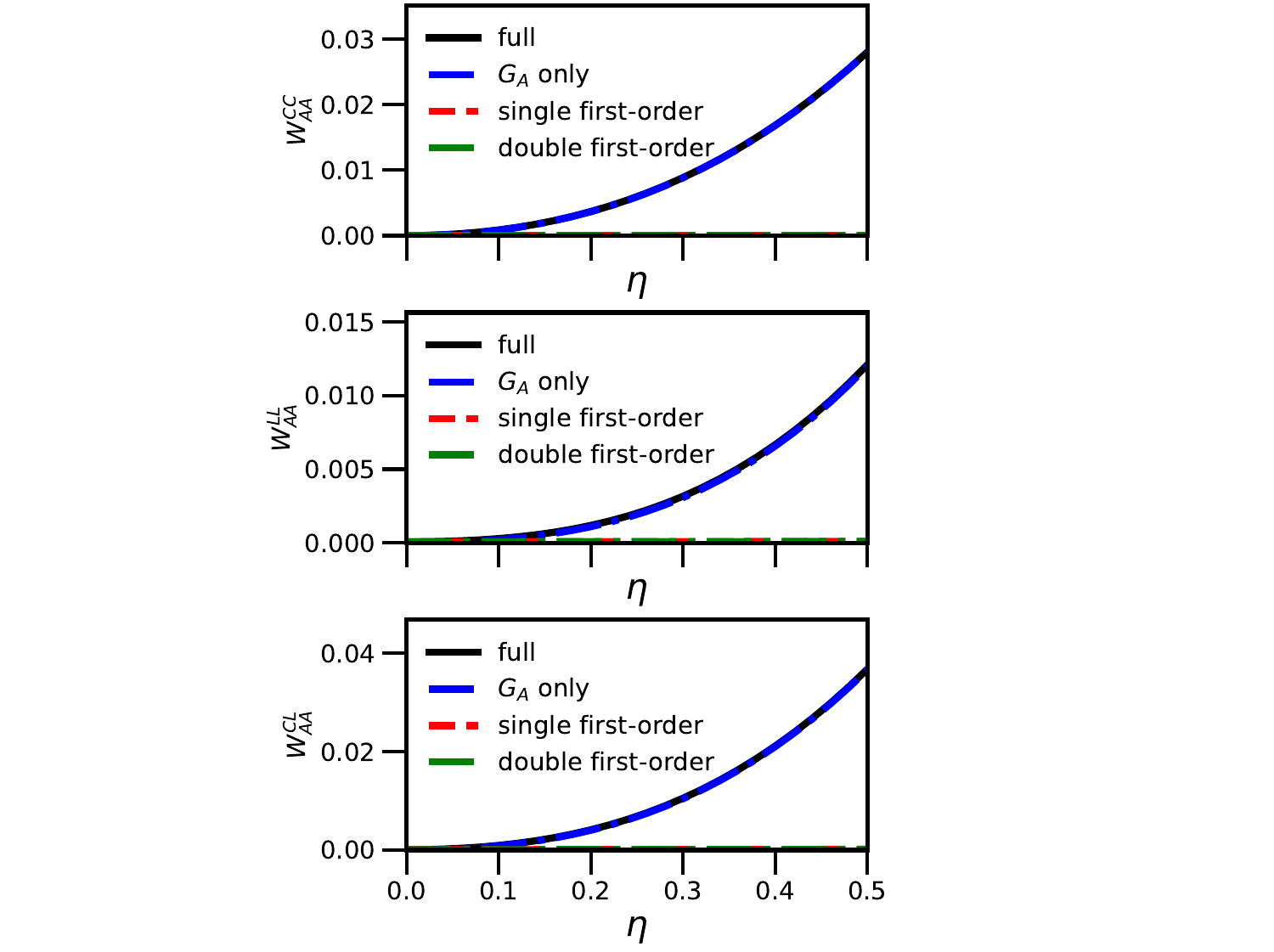}			
	\caption{The AA responses are shown as functions of $\eta$ for $\tau = 0.25$ and
		$\theta = 45^o$. The top panel shows $W^{CC}_{AA}$, the middle panel shows $ W^{LL}_{AA}$, and the bottom panel shows
		$W^{CL}_{AA}$.
		The solid line in these four panels shows the full solution, the red dashed line the single first-order approximation and the dash-dotted green line the double first-order approximation. The blue dash-dotted shows the contribution with just the $G_A$ terms. }
	\label{fig-respaatheta45}
\end{figure}

We start out with numerical results for the AA responses which have both $G_A$ and $G_A'$ terms, namely
$W^{CC}_{AA}, W^{LL}_{AA}$ and $W^{CL}_{AA}$, shown in Fig. \ref{fig-respaatheta45}. 
Besides the full, single first-order and
double first-order results discussed above,  we also 
show the ``$G_A$ only'' result, {\it i.e.,} the result with only the piece of the response proportional to $G_A$, with the term including $G_A'$ dropped. Looking at the results for $W^{CC}_{AA}, W^{LL}_{AA}$ and $W^{CL}_{AA}$, one sees that even the full result is very small,
as the terms lowest order in $\delta$, which would, in principle, be dominant, are multiplied with the small $G_A'^2$, making the low-$\eta$ expansion a moot point. The single first-order and double first-order approximations are numerically negligible compared
to the full result as they are proportional to $G_A'^2$. The tiny numerical difference between single and double first-order expressions
is irrelevant here. The dominant contribution to the four responses shown is the term $\propto G_A^2$, even though
it is multiplied by $\delta^2$.

\begin{figure}
	\centering
	\includegraphics[width=18cm]{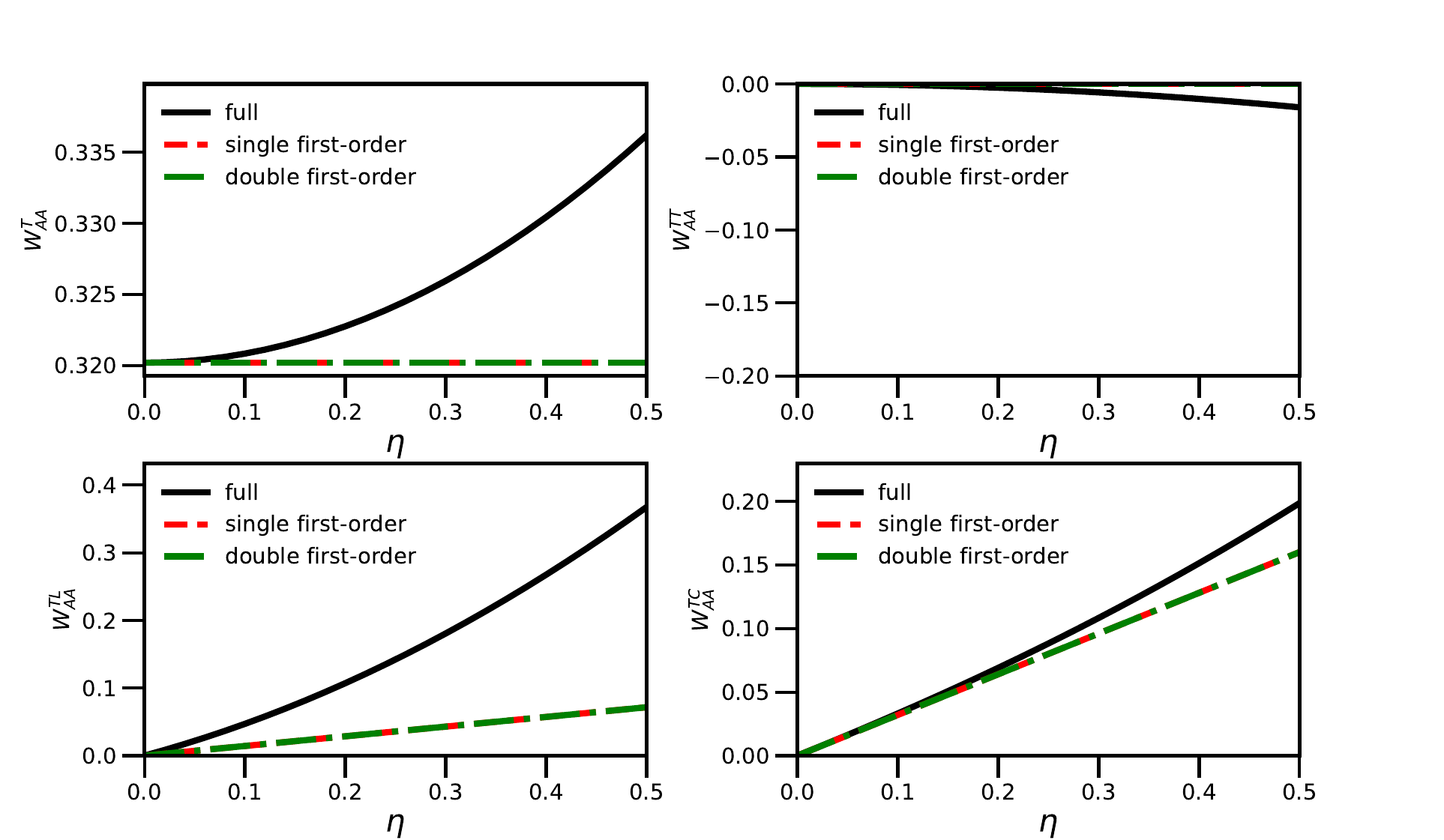}				
	\caption{The AA responses are shown as functions of $\eta$ for $\tau = 0.25$ and
		$\theta = 45^o$ and $\phi = 0^o$. The top row shows $W^{T}_{AA}$ (left) and  $W^{TT}_{AA}$ (right), and the bottom row
		 shows $ W^{TL}_{AA}$ (left) and $W^{TC}_{AA}$ (right).
		The solid line shows the full solution, the red dashed line the single first-order approximation and the dash-dotted green line the double first-order approximation. The contribution with only the $G_A$ terms coincides with the full solution and is not shown. }
	\label{fig-respintaatheta45}
\end{figure}

In Fig. \ref{fig-respintaatheta45} we show the responses $W^T_{AA}, W^{TT}_{AA}, W^{TL}_{AA}$ and $W^{TC}_{AA}$. None of these contains $G_A'$, 
and thus the leading-in-$G_A$ result coincides with the full result and is therefore not shown. The low-$\eta$ approximations now work as expected for these responses. 
As $W^T_{AA}$ (top left panel) does
not contain any explicit factors of $\kappa$ or $\lambda$, single first-order and double first-order terms coincide. 
The first-order approximations work quite well here for
all $\eta$ values shown. 
The $W^{TT}_{AA}$ response is purely ${\cal O}(\eta^2)$, so the first-order approximations for it vanish. The two other responses are of first-order in $\eta$, with single and double first-order coinciding. While the 
first-order approximations are very good at low-$\eta$, up to $\eta \approx 0.2$ for $W^{TC}_{AA}$ and up to $\eta \approx 0.1$ for $W^{TL}_{AA}$, they are not good at all for large values of $\eta > 0.4$ for $W^{TL}_{AA}$. For $W^{TC}_{AA}$, however, the first-order approximation is reasonable even up to $\eta \approx 0.5$.

\begin{figure}
	\centering
	\includegraphics[width=18cm]{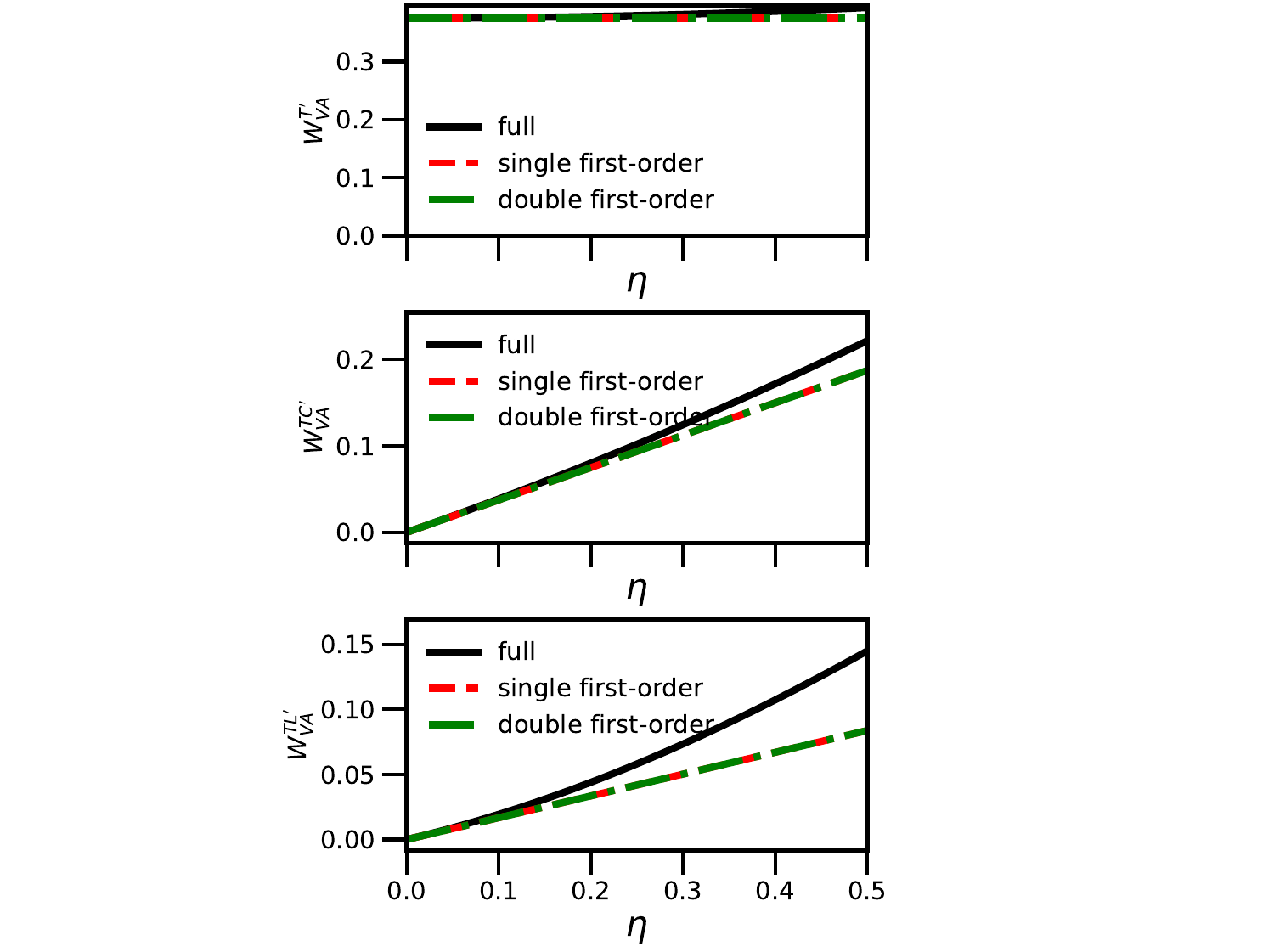} 			
	\caption{The VA responses are shown as functions of $\eta$ for $\tau = 0.25$ and
		$\theta = 45^o$ and $\phi = 90^o$. The top shows $W^{T'}_{VA}$, the middle panel shows $ W^{TC'}_{VA}$, 
		and the bottom panel shows
		$W^{TL'}_{VA}$.
		The solid line shows the full solution, the red dashed line the single first-order approximation and the dash-dotted green line the double first-order approximation. The contribution with only the $G_A$ terms coincides with the full solution and is not shown.
	}
	\label{fig-respintvatheta45}
\end{figure}

Finally, we present the VA responses in Fig. \ref{fig-respintvatheta45}. For these three responses, once more the smaller form factor $G_A'$ does not show up, 
and thus the leading-in-$G_A$ result coincides with the full result and is therefore not shown separately.
For all three responses, single first-order and double-first order
approximations are identical.  $W^{T'}_{VA}$ is the only response non-zero in the rest system, and the first-order approximation is excellent up to $\eta \approx 0.2$ and still very good for all $\eta$ considered here.
For $ W^{TC'}_{VA}$, the first-order approximation shows the same behavior, {\it i.e.,} it is excellent up to $\eta \approx 0.2$ and still very good for all $\eta$ considered here. For $W^{TL'}_{VA}$, the quality of the first-order approximation is fine up to $\eta \approx 0.1$, but then it quickly deviates from the full result.

\begin{figure}
	\centering
	\includegraphics[width=18cm]{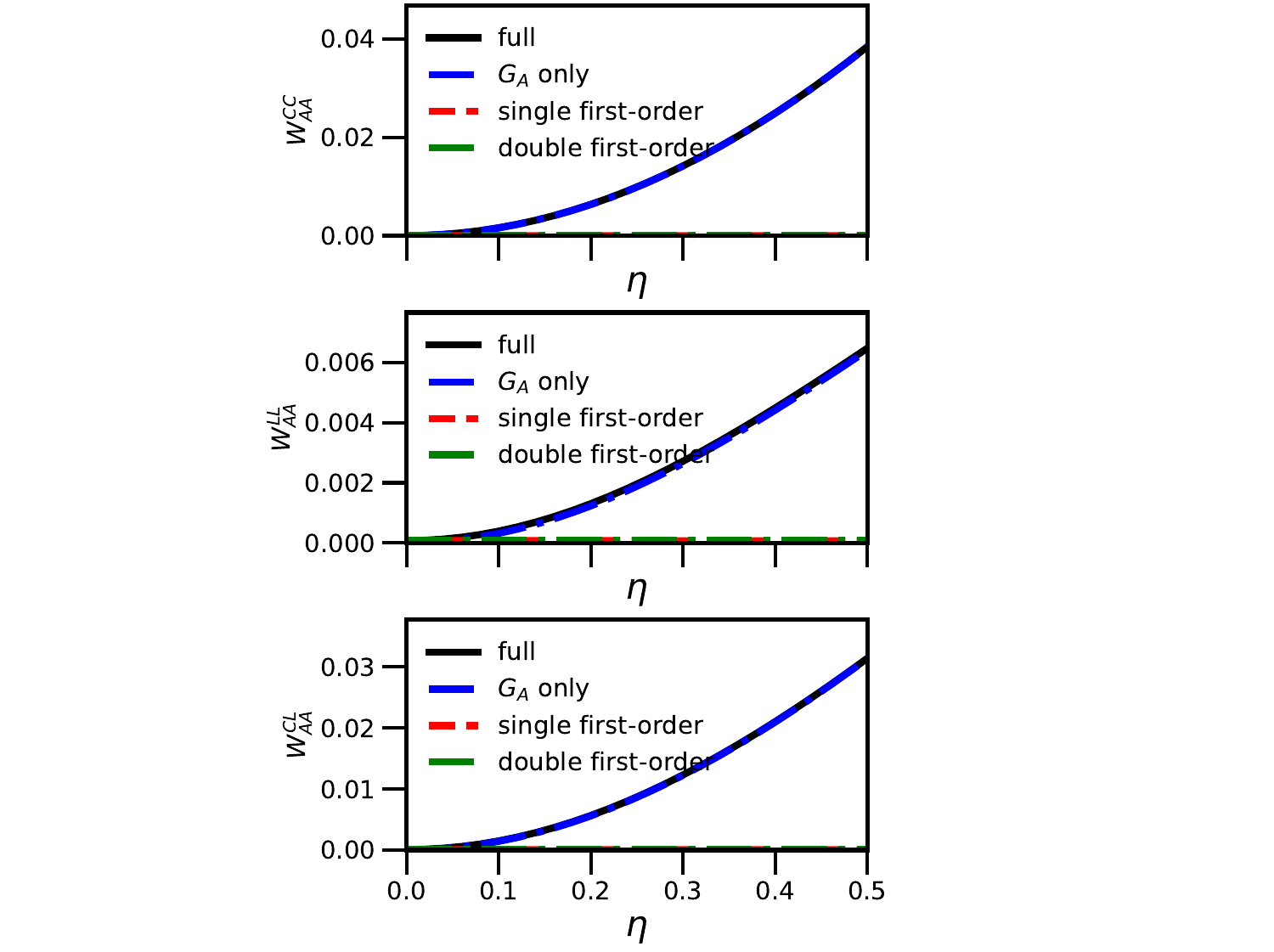} 			
	\caption{The AA responses are shown as functions of $\eta$ for $\tau = 0.25$ and
		$\theta = 45^o$. The top panel shows $W^{CC}_{AA}$, the middle panel shows $ W^{LL}_{AA}$, and the bottom panel shows
		$W^{CL}_{AA}$.
		The solid line in these four panels shows the full solution, the red dashed line the single first-order approximation and the dash-dotted green line the double first-order approximation. The blue dash-dotted shows the contribution with just the $G_A$ terms. }
	\label{fig-respaatheta90}
\end{figure}
\begin{figure}
	\centering
	\includegraphics[width=18cm]{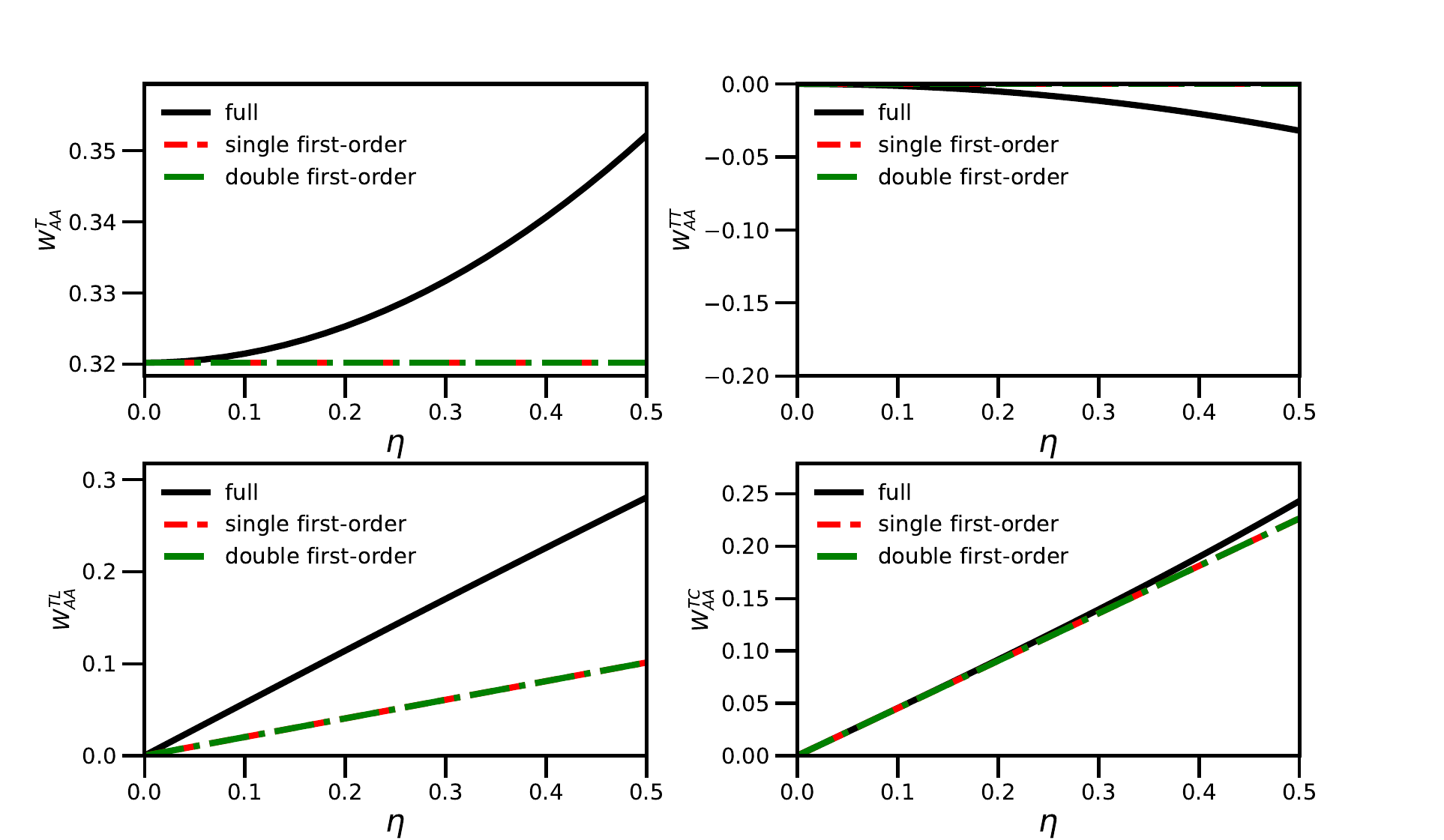} 	 				
	\caption{The AA responses are shown as functions of $\eta$ for $\tau = 0.25$ and
		$\theta = 90^o$ and $\phi = 0^o$. The top row shows $W^{T}_{AA}$ (left) and  $W^{TT}_{AA}$ (right), and the bottom row
		shows $ W^{TL}_{AA}$ (left) and $W^{TC}_{AA}$ (right).
		The solid line shows the full solution, the red dashed line the single first-order approximation and the dash-dotted green line the double first-order approximation. The contribution with just the $G_A$ terms coincides with the full solution and is not shown. }
	\label{fig-respintaatheta90}
\end{figure}
\begin{figure}
	\centering
	\includegraphics[width=18cm]{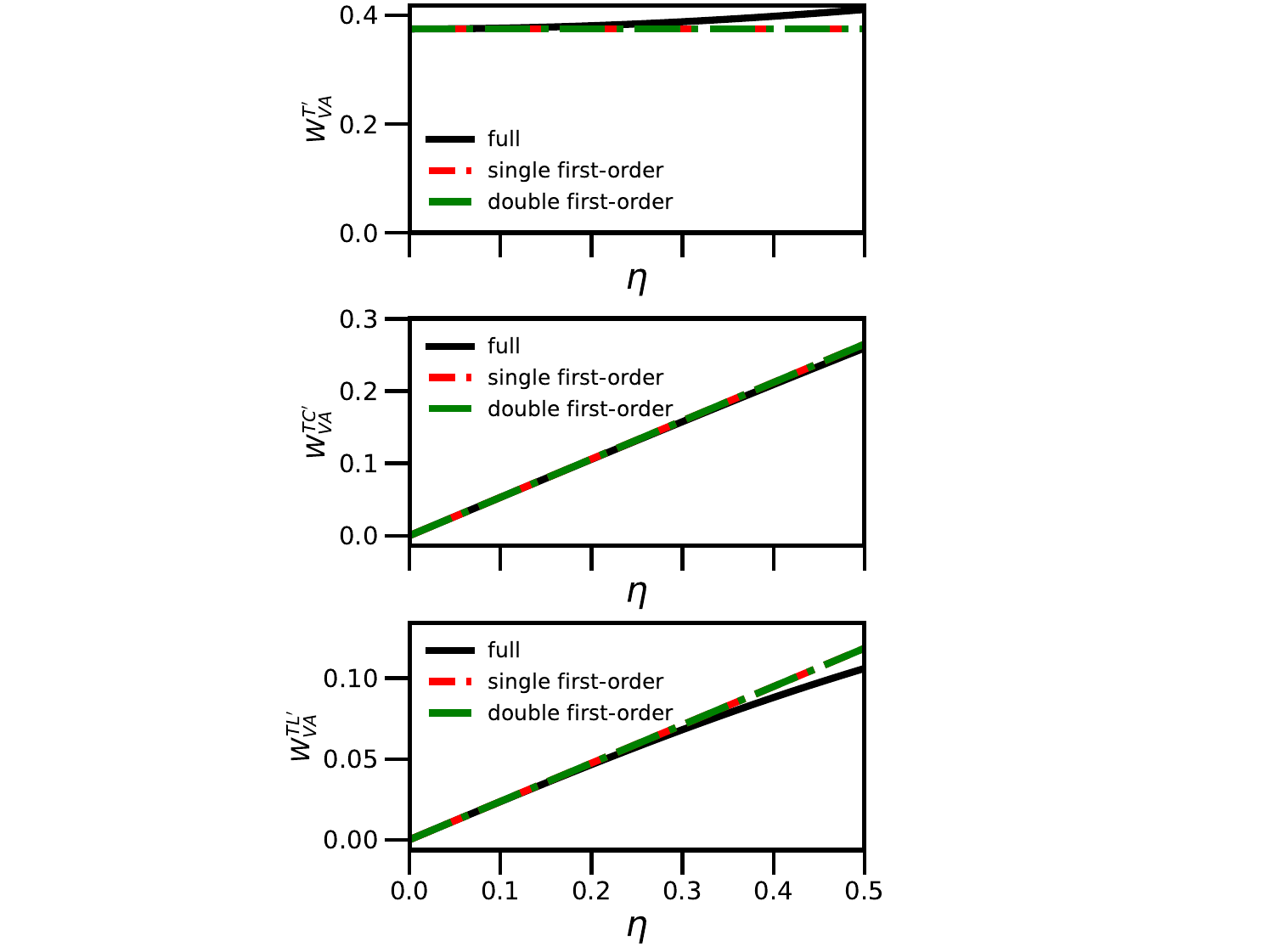}			
	\caption{The VA responses are shown as functions of $\eta$ for $\tau = 0.25$ and
		$\theta = 90^o$ and $\phi = 90^o$. The top shows $W^{T'}_{VA}$, the middle panel shows $ W^{TC'}_{VA}$, 
		and the bottom panel shows
		$W^{TL'}_{VA}$.
		The solid line shows the full solution, the red dashed line the single first-order approximation and the dash-dotted green line the double first-order approximation. The contribution with just the $G_A$ terms coincides with the full solution and is not shown.
	}
	\label{fig-respintvatheta90}
\end{figure}

In Fig. \ref{fig-respaatheta90} we show the same responses as in Fig. \ref{fig-respaatheta45}, but for the angle $\theta = 90^o$. 
All the qualitative features of the results for $\theta = 45^o$ appear here, too. Since $\delta = \eta \sin \theta$ is larger for the same value
of $\eta$ for $\theta = 90^o$, the difference between the full result and the single and double first-order results for $W^T_{AA}$
(see Fig. 7)
is slightly more pronounced at larger $\eta$. For $W^{LL}_{AA}$, the difference between the full and leading-in-$G_A$ result is more
pronounced for the same reason, even though these two curves still almost coincide.

In Fig. \ref{fig-respintaatheta90} we show the four responses $W^{T}_{AA}, W^{TT}_{AA}, W^{TL}_{AA}$ 
and $W^{TC}_{AA}$ for $\theta = 90^o$, the same responses as in Fig. 
\ref{fig-respintaatheta45}. Once more the results for $\theta = 45^o$ and $\theta = 90^o$ are qualitatively very similar. The most
noticeable difference is that for $W^{TC}_{AA}$ and $W^{TL}_{AA}$, the first-order approximations now are closer to the full
result. This stems from the behavior of $\bar{\epsilon}$ which varies much less for $\theta = 90^o$ than $\theta = 45^o$
(see Fig. 4 in \cite{Donnelly:2024hqx}), and is a multiplicative factor in both responses.

In Fig. \ref{fig-respintvatheta90} we show responses $W^{T'}_{VA}, W^{TC'}_{VA}$ and $W^{TL'}_{VA}$, the same responses as in Fig. 
\ref{fig-respintvatheta45}, but for the angle $\theta = 90^o$. . Just as for the previous figure, the responses are qualitatively very similar for the two values of $\theta$.

\begin{figure}
	\centering
	\includegraphics[width=18cm]{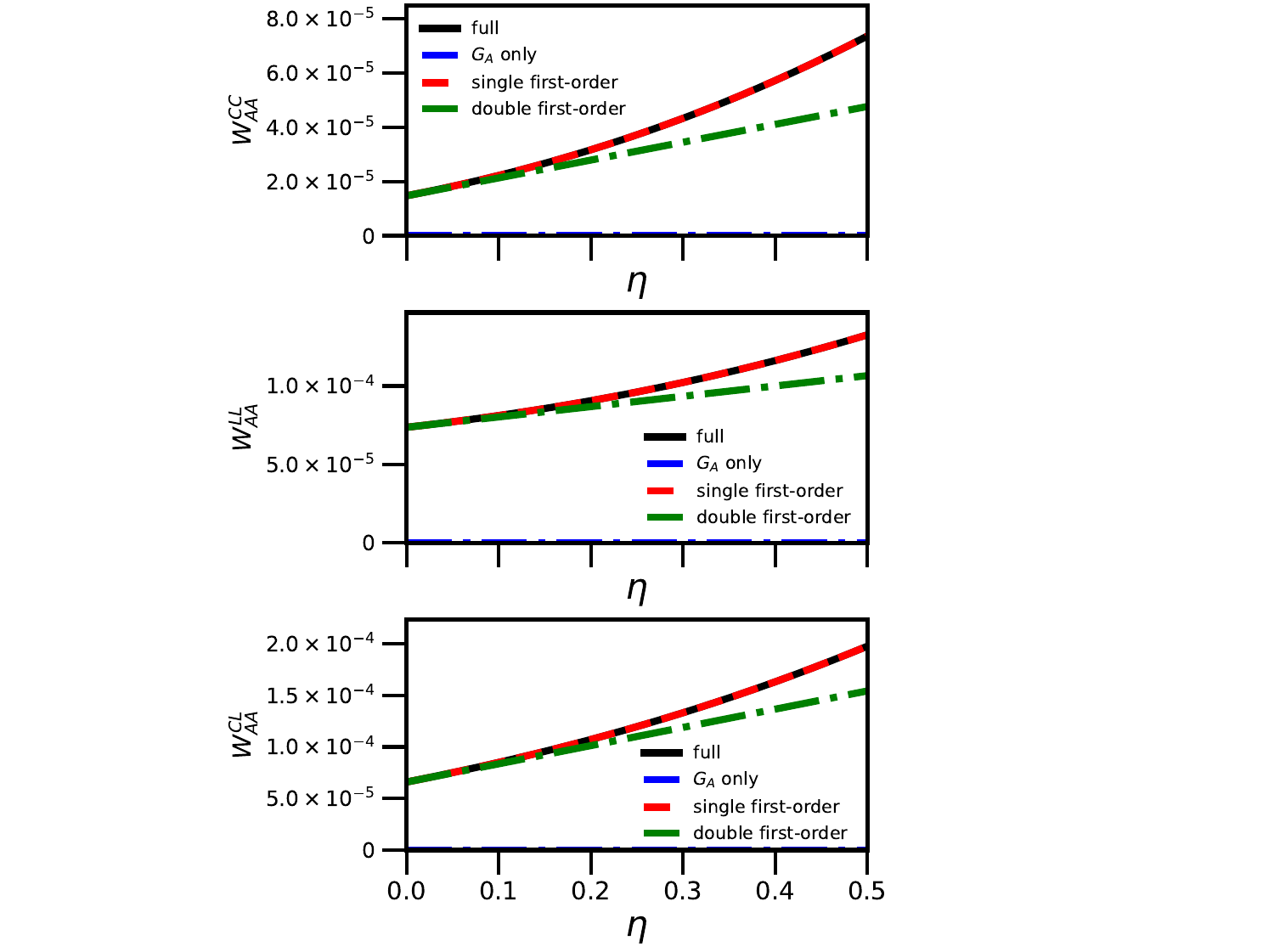} 				
	\caption{ The AA responses are shown as functions of $\eta$ for $\tau = 0.25$ and
		$\theta = 0^o$. The top panel shows $W^{CC}_{AA}$, the middle panel shows $ W^{LL}_{AA}$, and the bottom panel shows
		$W^{CL}_{AA}$.
		The solid line in these four panels shows the full solution, the red dashed line the single first-order approximation and the dash-dotted green line the double first-order approximation. The blue dash-dotted shows the contribution with just the $G_A$ terms. }
	\label{fig-respaatheta0}
\end{figure}

For $\theta = 0^o$ and $\theta = 180^o$, $\delta = 0$ and five of our responses then vanish: $W^{TT}_{AA} = W^{TC}_{AA} = 
W^{TL}_{AA} = W^{TC'}_{VA} = W^{TL'}_{VA} = 0$. Two of the responses become independent of $\eta$:  $W^T_{AA}$  reduces to $ 2 (1 + \tau) G_A^2$ for $\delta = 0$, and $W^{T'}_{VA}$ takes the form $ 2 \sqrt{\tau (1+\tau)} G_M G_A$ in this case.  Only $W^{CC}_{AA}, W^{CL}_{AA}$ and $W^{LL}_{AA}$ show a non-trivial dependence on $\eta$ for $\delta = 0$. These responses are shown in Fig. \ref{fig-respaatheta0}. Now, the $G_A$ only contributions to these three responses are zero, as $G_A$ is multiplied with $\delta^2$ for these responses. The single first-order terms now coincide with the full result, while the double first-order approximation is valid only for $ \eta \leq 0.2$.

For $\theta = 180^o$, the responses behave qualitatively in the same manner,
but are noticeably smaller than for $\theta = 0^o$.

 We now look at the rather small value of $\tau = 0.005$, for which the ratio of $G_A/G_A' \approx 2.4$. This means that for these
kinematics, terms including $G_A'$ are not automatically much smaller than all others, and we can observe more interesting structures
in the responses at these low-$\tau$ kinematics.

In Fig. \ref{fig-kinematicsall}, we show the kinematics variables for fixed $\tau = 0.005$ and $\theta = 0^o$ (top),
 $\theta = 45^o$ (middle) and $\theta = 90^o$ (bottom)
as functions of $\eta$. It will be relevant later on that the values of
$\lambda$, which are small at the smaller values of $\theta$, are especially small with growing $\eta$, approaching zero.
We note in passing  that for larger $\theta$ values than shown, the physical region is limited to small $\eta$
values, {\it e.g.} for  $\theta = 110^o$,
the physical region extends only to values of $\eta$ smaller than $~0.2$.

	\begin{figure}
		\centering
		\includegraphics[width=18cm]{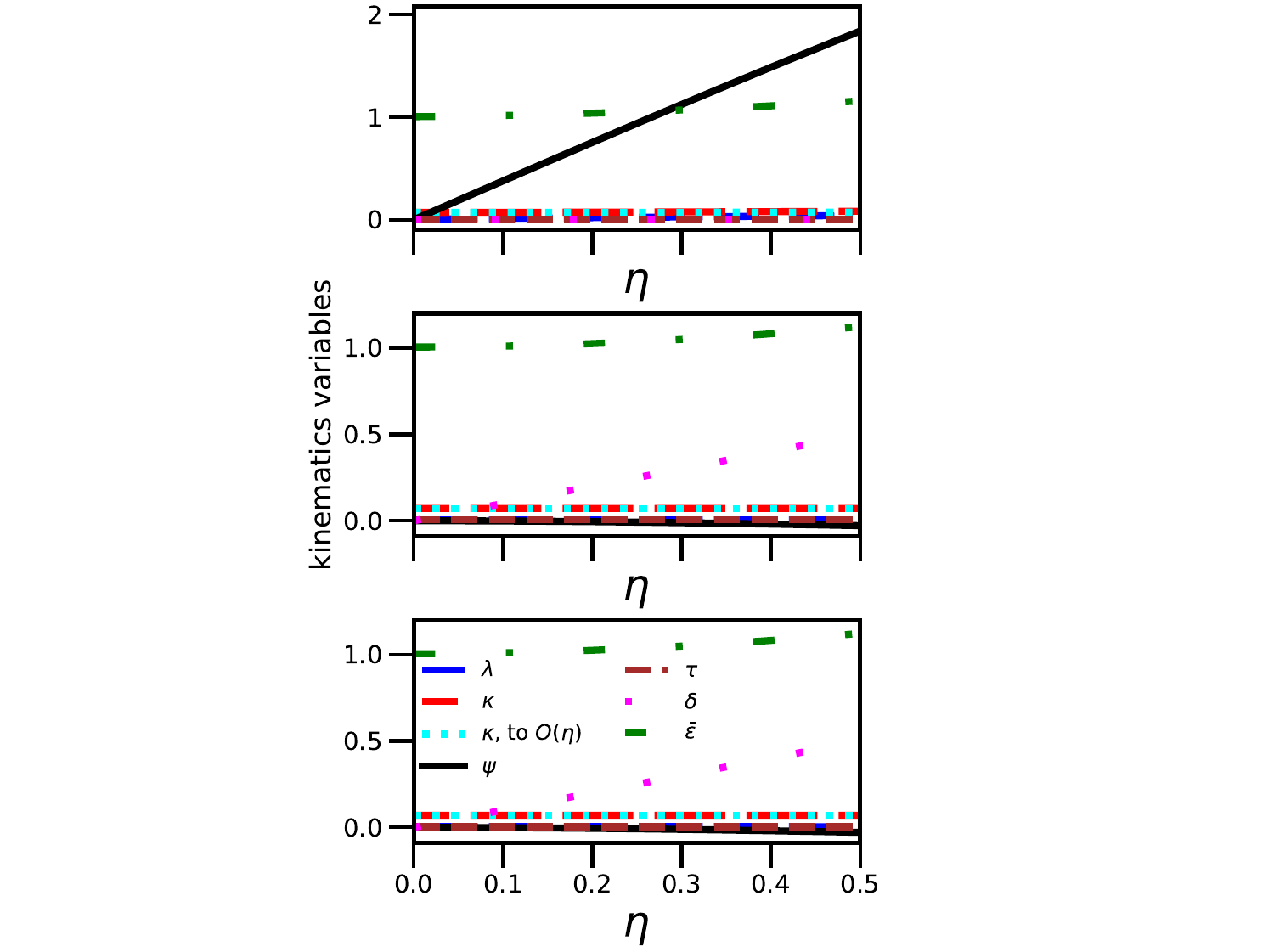} 
		\caption{Various dimensionless kinematic variables are shown as functions of $\eta$ for $\tau = 0.005$ and
			$\theta = 0^\circ$ (top panel),
			$\theta = 45^\circ$ (middle panel), and $\theta = 90^\circ$ (bottom panel). For each panel,
			the black solid line is $\psi$, the blue long-dash-dotted line is $\lambda$, the red long-dashed line is $\kappa$, the cyan dotted line is $\kappa$ to $O(\eta)$, the brown dashed line is $\tau$,
			the red loosely dotted line is $\delta$, and the green loosely dash-dotted line is $\bar{\epsilon}$.
		}
		\label{fig-kinematicsall}
	\end{figure}

\begin{figure}
		\centering
		\includegraphics[width=18cm]{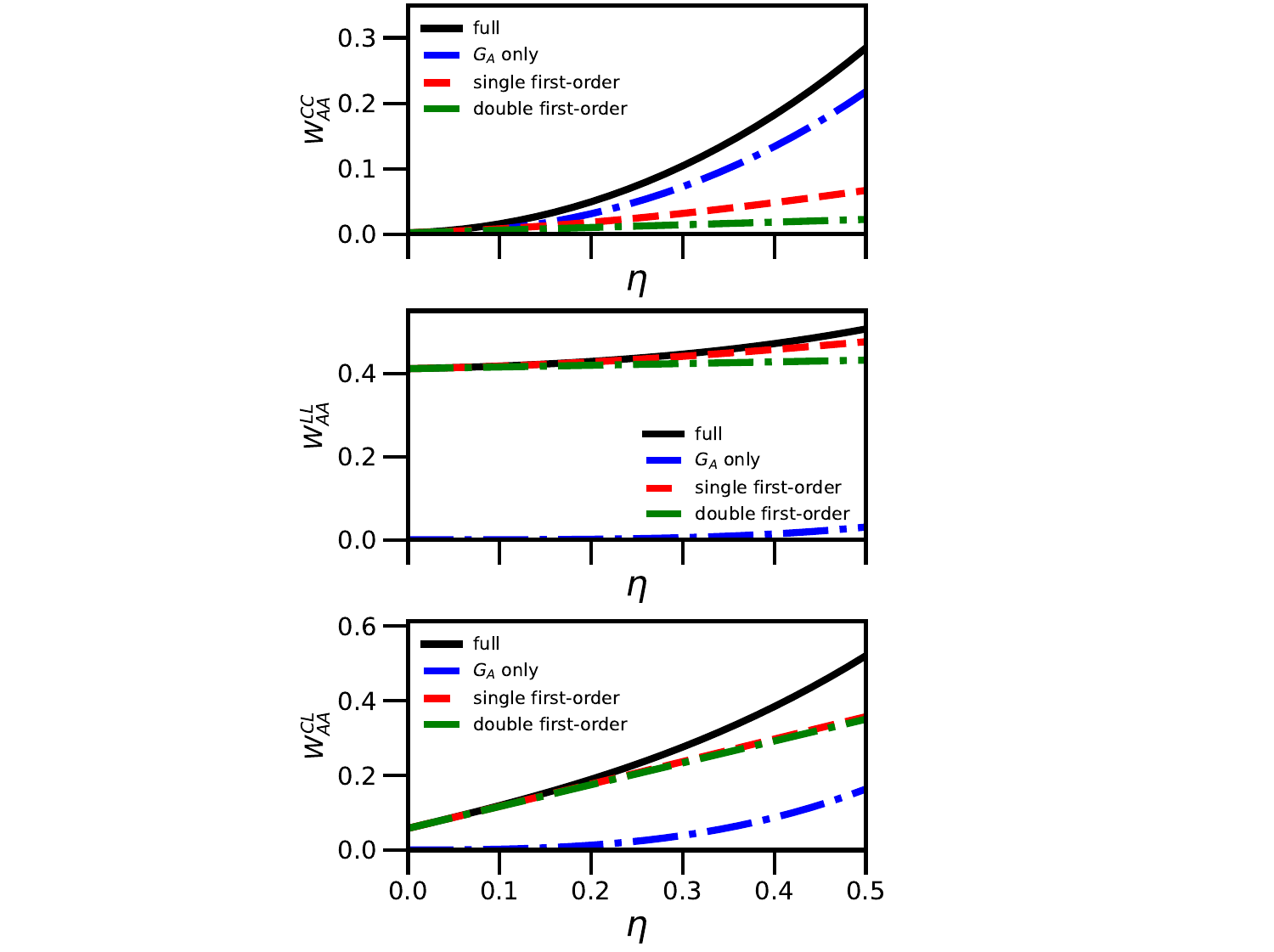} 				
		\caption{The AA responses are shown as functions of $\eta$ for $\tau = 0.005$ and
			$\theta = 45^o$. The top panel shows $W^{CC}_{AA}$, the middle panel shows $ W^{LL}_{AA}$, and the bottom panel shows
			$W^{CL}_{AA}$.
			The solid line in these four panels shows the full solution, the red dashed line the single first-order approximation and the dash-dotted green line the double first-order approximation. The blue dash-dotted shows the contribution with just the $G_A$ terms. }
		\label{fig-respaatheta45tausmall}
	\end{figure}

	Now we have a noticeable contribution from terms $\propto G_A'$ for the first time as shown in Fig. \ref{fig-respaatheta45tausmall}. The responses
	$W^{CC}_{AA}, W^{LL}_{AA}$ and $W^{CL}_{AA}$ now have a distinct difference between the $G_A$-only and full result. This difference is particularly pronounced for the two latter responses.
	In the expression for $W^{LL}_{AA}$ , $G_A^{'2}$ appears by itself, whereas $G_A$ is multiplied not just by $\delta^2$, but also by the factor $(\lambda/\kappa)^2$, which is less than one
	for spacelike kinematics. The contribution with $G_A$-only is thus close to zero for $W^{LL}_{AA}$. For $W^{CL}_{AA}$, the relative factor between the two different form factors is $\delta^2$,
	leading to a very distinct difference in the numerical results for $G_A$-only and the full calculation. This difference is still noticeable, but much smaller for $W^{CC}_{AA}$ as here,
	$G_A^{'2}$ is also multiplied by the factor $(\lambda/\kappa)^2$, thus washing out the difference a bit. Now that the first-order terms, which are all proportional to $G_A^{'2}$, are actually numerically relevant, 
	one can see that both first-order approximations work quite nicely up to $ \eta \approx 0.2$ for $W^{LL}_{AA}$ and $W^{CL}_{AA}$, and up to $\eta \approx 0.1$ for $W^{CC}_{AA}$.
	For $W^{LL}_{AA}$, the single first-order approximation continues to work well up to the highest $\eta$ values shown, and the double first-order results are still fairly close to the full result at these values. 
	This is in contrast to the behavior for $W^{CC}_{AA}$ and $W^{CL}_{AA}$, where the quality of the first-order approximations rapidly deteriorates for medium and higher $\eta$ values.
	
	We only show numerical results for $W^{CC}_{AA}, W^{LL}_{AA}$ and $W^{CL}_{AA}$ at small $\tau$, as all the other responses 
	do not have any contributions from $G_A'$, and thus their behavior for small $\tau = 0.005$ is extremely similar
	to their behavior at large $\tau = 0.25$, shown above. The point of looking at $\tau  = 0.005$ is to study numerically
	relevant contributions from the $G_A'$ terms.

\begin{figure}
		\centering
		\includegraphics[width=16cm]{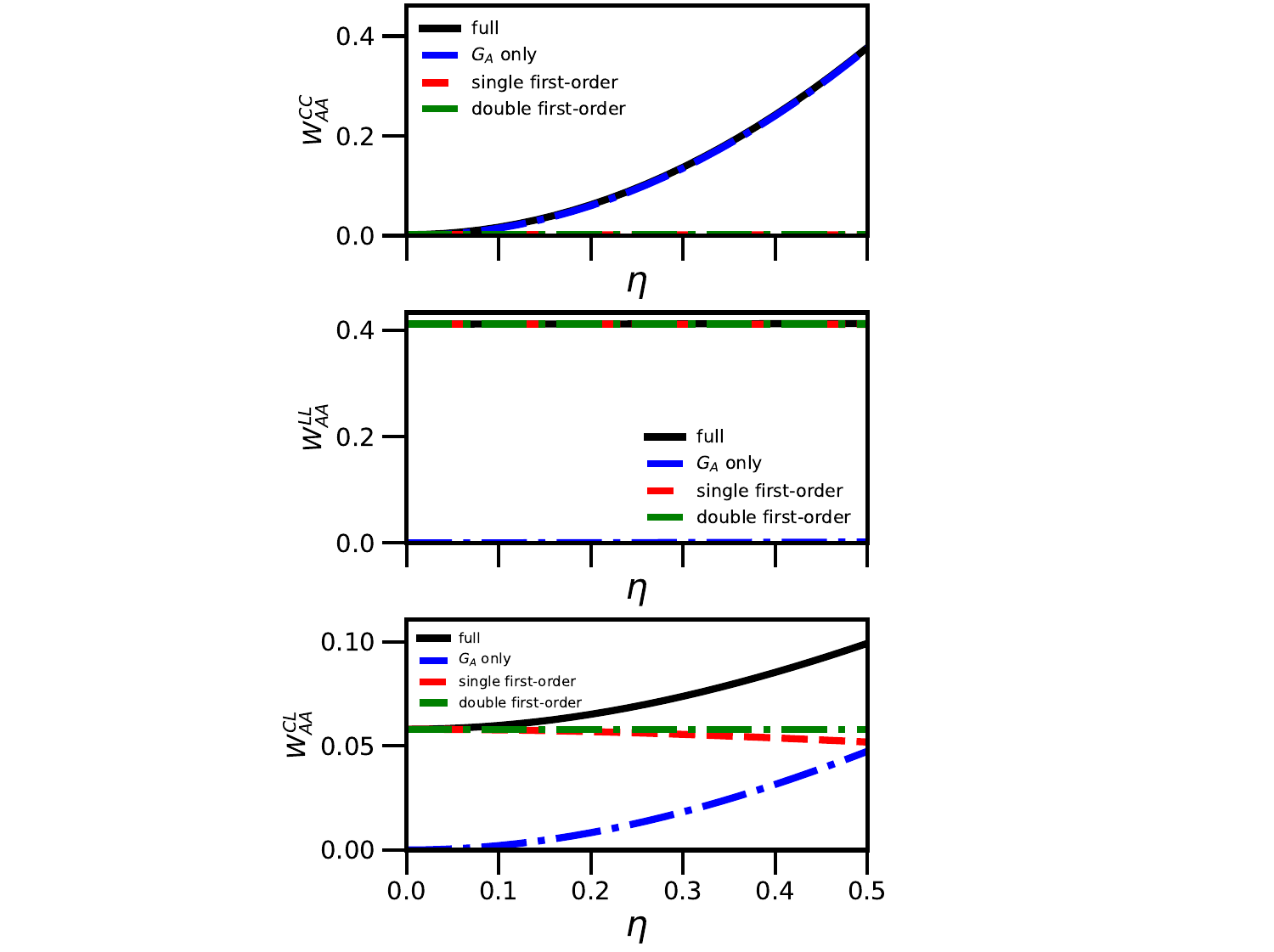} 				
		\caption{The AA responses are shown as functions of $\eta$ for $\tau = 0.005$ and
			$\theta = 90^o$.  The top panel shows $W^{CC}_{AA}$, the middle panel shows $ W^{LL}_{AA}$, and the bottom panel shows
			$W^{CL}_{AA}$.   
			The solid line in these four panels shows the full solution, the red dashed line the single first-order approximation and the dash-dotted green line the double first-order approximation. The blue dash-dotted shows the contribution with just the $G_A$ terms. }
		\label{fig-respaatheta90tausmall}
	\end{figure}

	In Fig. \ref{fig-respaatheta90tausmall} we now show the same responses as in Fig. \ref{fig-respaatheta45tausmall}, but for the angle $\theta = 90^o$. The results for $\theta = 90^o$ are quite similar to the results for $\theta = 45^o$.
	The main difference between these two angles is driven by the fact that
	for the $90^o$ kinematics, the value of $\lambda$ is extremely small for all $\eta$ values. Due to this, the contribution to $W^{CC}_{AA}$ from the terms with $G_A'$, {\it i.e.,} the first-order contribution, becomes negligible, and the $G_A$-only
	results are almost the same as the full result, thus resembling the situation for larger $\tau$. For $W^{LL}_{AA}$, where the $G_A$ term is multiplied with $(\lambda/\kappa)^2$, the effect is the opposite: the $G_A$-only
	results are close to zero, and the two first-order approximations are practically perfect for all $\eta$ values shown. For $W^{CL}_{AA}$,  $\lambda/\kappa$ is a common factor for both $G_A^2$ and $G_A^{'2}$ terms, and
	so the overall magnitude of the response is reduced by roughly a factor of $3$ compared to $\theta = 45^o$, with much smaller changes in overall size for the other three responses shown. The relative behavior of
	the full, first-order and $G_A$-only results is quite similar. The most notable difference is that the single first-order and double first-order results deviate a bit more from each other, and remain mostly flat.
	Once again, the first-order approximations are good only up to $\eta \approx 0.1$.

	

	For $\theta$ values larger than $90^o$, we run out of phase space for our small $\tau$ value, {\it e.g.,} for $\theta = 110^o$, the allowed kinematics region  extends only up to $\eta \approx 0.2$.


\section{\protect\bigskip Conclusions \label{sec-conclusions}}

In this study we have provided the covariant, axial-vector single-nucleon current matrix elements restricted only by limiting the focus to first-class operators. This constitutes a natural extension of our previous studies of the covariant, vector EM current matrix elements in \cite{JandD,Donnelly:2024hqx}. As in that previous work we have two motivations; (1) to yield the matrix elements needed to give the cross sections for electroweak reactions with single nucleons when the latter are moving (in contrast to the usual situation where the target rest frame is typically assumed), and (2) to organize the results in forms where convenient approximations can be made. With respect to point (1), in \cite{Donnelly:2024hqx} we studied electron scattering from nucleons and in fact assumed that the struck (moving) nucleon was polarized, as was the incident electron. Here in the present work we have not assumed that the incident leptons and nucleons are polarized, but have include both vector and axial-vector contributions in the analysis. With respect to point (2), we have set up the formalism to facilitate the implementation of the ``standard prescription for nuclear physics". Typically one studies the free (on-shell) lepton-nucleon problem and then carries what arises over to the problem of nucleons within nuclei by replacing the 3-momentum of the incident nucleon ${\mathbf p}$ with $-i{\mathbf \nabla}$ and working in coordinate space. In the present study we have organized the formalism in a way that allows general values for the energy, 3-momentum and 4-momentum transfers, $\omega$, ${\mathbf q}$ and $Q^2 $, respectively, while isolating the dependences on ${\mathbf p}$ so that various approximations via expansions in small ${\mathbf \eta} \equiv {\mathbf p}/m_N$, where $m_N$ is the nucleon mass, can be made.


Instead of employing the axial-vector $G_A$ and induced pseudoscalar $G_P$ form factors in the definition of the general --- but for the restriction to first-class currents --- axial-vector current matrix element, we have used in the present work the form factor $G_A$ together with $G'_A \equiv G_A - \tau G_P$, where $\tau \equiv |Q^2|/4 m_N^2$ . This has a similar effect to that which occurs when defining the (vector) Sachs form factors in that it eliminates terms that go as $G_A G'_A$. Moreover, upon assuming a model for $G_P$ which invokes pion pole dominance, one finds that, except at very low values of $\tau$, one has $|G'_A/G_A| \ll 1$, which leads to a very good approximation, namely dropping all contributions involving $G'_A$ except at very small $\tau$.

We have also followed the ideas presented in \cite{Donnelly:2024hqx} and obtained the unpolarized isovector response functions that occur in studies of charge-changing neutrino and anti-neutrino reactions, namely, for the reactions  $\nu_\ell + n \rightarrow \ell^- + p$ and ${\bar\nu}_\ell + p \rightarrow \ell^+ + n$, where $\ell = e$, $\mu$ or $\tau$. We have again organized the expressions for the response functions in forms where approximations coming from expansions in ${\mathbf \eta}$ can conveniently be undertaken. Results are also obtained for rest-frame conditions and for the Breit frame.

Finally, we have used the insights gotten in \cite{Donnelly:2024hqx} to obtain numerical results for the various response functions in typical conditions to help in assessing the quality of the different approximation schemes.

Our numerical results show that for the three responses with both form factors $G_A$ and $G_A'$, namely $W^{CC}_{AA}, W^{CL}_{AA}$
and $W^{LL}_{AA}$, using just the terms with $G_A$ and neglecting all $G_A'$ terms is an excellent approach at values of $\tau \geq 0.01$.
In these kinematics, the expansion in order of $\eta$ becomes meaningless, as $G_A$ terms dominate the numerical results.

For very small values of $\tau$, where $G_A'$ is at least somewhat comparable to $G_A$, the expansion in $\eta$ becomes more meaningful.
Our results show that depending on the specific response, the first-order approximations can be valid up to values of $\eta \approx 0.2$.

For the transverse responses $W^T_{AA}, W^{TT}_{AA}, W^{TL}_{AA}$ and $W^{TC}_{AA}$, only $G_A$ enters, and the expansion in $\eta$ is more
natural. The single first-order and double first-order approximations coincide. Our numerical results have shown that the first-order
approximations are valid at least up to $\eta \approx 0.1$, and at most up to $\eta \approx 0.5$, depending strongly on the response in question
and also on the value of $\theta$. The situation is similar for the $VA$ responses.
Thus, we conclude that when using a first-order approximation, it is best to be careful and check the specific response and kinematics,
	as the quality of these approximations varies greatly.

\appendix

\section{Conventions\label{sec-conventions}}

In this work we employ the conventions adopted in previous work including \cite{Donnelly:1985ry,Raskin:1988kc,Donnelly:2023rej,arxivlong}: 4-vectors are written $%
A^{\mu }=(A^{0},A^{1},A^{2},A^{3})=(A^{0},\mathbf{a})$ with capital letters
for the 4-vectors and lower-case letters for 3-vectors. The magnitude of a
3-vector is written as $a=|\mathbf{a}|$. One also has $A_{\mu }=g_{\mu \nu
}A^{\mu }=(A^{0},-A^{1},-A^{2},-A^{3})$ with%
\begin{equation}
g_{\mu \nu }=g^{\mu \nu }=\left( 
\begin{array}{llll}
1 & 0 & 0 & 0 \\ 
0 & -1 & 0 & 0 \\ 
0 & 0 & -1 & 0 \\ 
0 & 0 & 0 & -1%
\end{array}%
\right) .  \label{eq-app-conv-1}
\end{equation}%
The scalar product of two 4-vectors is given by $A\cdot B=A_{\mu }B^{\mu
}=(A^{0})^{2}-a^{2}$, following the conventions of \cite{Bjorken:1965sts}. For instance,
for the 4-momentum of an on-shell particle of mass $M$, energy $E_p$ and
3-momentum $p$ we have $P^{\mu }=(E_p ,\mathbf{p})$ and hence $P^{2}=P_{\mu
}P^{\mu }=E_p^{2}-p^{2}=M^{2}$. One problem occurs with these conventions, 
\textit{viz.} for the momentum transfer 4-vector we have $%
Q^{2}=(Q^{0})^{2}-q^{2}$ which, for electron scattering is spacelike, and
accordingly $Q^{2}<0$. One should be careful not to confuse our sign convention for this quantity with the so-called SLAC convention which has the opposite sign. The totally anti-symmetric Levi-Civita symbol follows the conventions
of \cite{Bjorken:1965sts} where%
\begin{equation}
\epsilon _{0123}=-\epsilon ^{0123}=+1. \label{eq-app-conv-2} 
\end{equation}
Also we take $\hbar = c =1$. When applying the Feynman rules we also employ the conventions of \cite{Bjorken:1965sts}.
In particular, following \cite{Bjorken:1965sts}, our spinors are normalized as $\bar{u} u = 1$.



{\bf Acknowledgments}: This work was
supported in part by funds provided by the National Science Foundation under
grant No. PHY-2208237 (S. J.).


\end{document}